\documentclass[nameyear,final]{econsocart}
\usepackage{amsmath}
\usepackage{amsfonts}
\usepackage{amssymb}
\usepackage{graphicx}
\usepackage{color}
\usepackage{rotating}
\usepackage{url}
\usepackage{bm}
\usepackage{longtable}
\usepackage{multirow} 

\RequirePackage[colorlinks,citecolor=blue,linkcolor=blue,urlcolor=blue,pagebackref]{hyperref}

\startlocaldefs

\DeclareFontFamily{U}{mathx}{}
\DeclareFontShape{U}{mathx}{m}{n}{<-> mathx10}{}
\DeclareSymbolFont{mathx}{U}{mathx}{m}{n}
\DeclareMathAccent{\widehat}{0}{mathx}{"70}
\DeclareMathAccent{\widecheck}{0}{mathx}{"71}

\theoremstyle{plain}
\newtheorem{theorem}{Theorem}

\newtheorem{proposition}{Proposition}

\newtheorem{assumption}{Assumption}
\theoremstyle{definition}
\newtheorem{definition}{Definition}

\endlocaldefs

\begin{document}

\begin{frontmatter}

\title{Efficient Mechanisms under Unawareness}
\runtitle{Efficient Mechanisms under Unawareness}

\begin{aug}
%
%
%
\author[add1]{\fnms{Kym}~\snm{Pram}\ead[label=e1]{kpram@unr.edu}}
\author[add2]{\fnms{Burkhard C.}~\snm{Schipper}\ead[label=e2]{bcschipper@ucdavis.edu}}
\address[add1]{%
\orgdiv{Department of Economics},
\orgname{University of Nevada, Reno}}

\address[add2]{%
\orgdiv{Department of Economics},
\orgname{University of California, Davis}}

\end{aug}

\begin{funding}
04/05/2025. We thank Sarah Auster, Botond Koeszegi, Takashi Kunimoto, Matthias Lang, Lu Jingfeng, Moritz Meyer-ter-Vehn, Benny Moldovanu, Alessandro Pavan, Tomasz Sadzik, Klaus Schmidt, Roland Strausz, Tymon Tatur, Joel Watson, and participants in seminars at Bonn, HU Berlin, UC Davis, UC Irvine, LMU Munich, NSU Singapore, SWET 2025, SEA 2023, and SAET 2023 for useful discussions. Burkhard gratefully acknowledges financial support via ARO Contract W911NF2210282.
\end{funding}
%

\begin{abstract}
We study the design of efficient mechanisms under asymmetric awareness and information. Unawareness refers to the lack of conception rather than the lack of information. Assuming quasi-linear utilities and private values, we show that we can implement in conditional dominant strategies a social choice function that is utilitarian ex-post efficient when pooling all awareness of all agents without the need of the social planner being fully aware ex-ante. To this end, we develop novel dynamic versions of Vickrey-Clarke-Groves mechanisms in which types are revealed and subsequently elaborated at endogenous higher awareness levels. We explore how asymmetric awareness affects budget balance and participation constraints. We show that ex-ante unforeseen contingencies are no excuse for deficits. Finally, we propose a modified reverse second price auction for efficient procurement of complex incompletely specified projects.
\end{abstract}


\end{frontmatter}

\section{Introduction}\label{s1}

Mechanism design studies the design of institutions governing collective decisions such as markets, contracts, or political systems in the presence of asymmetric information. However, agents may not just face asymmetric information but also asymmetric awareness. Unawareness refers to the lack of conception rather than the lack of information. Agents and the designer of mechanisms may be unaware of some events and actions affecting values and costs of complex private or public projects, allocations, and outcomes and may not even realize this fact. In this paper, we extend mechanism design to unawareness. Under the assumption of quasi-linear preferences, we show how to design efficient mechanisms in the presence of asymmetric awareness. To this end, we introduce dynamic direct elaboration mechanisms in which not only are types communicated from agents but also awareness is raised among participants in back-and-forth communication between participants and transfers are inspired by Vickrey-Clarke-Groves (VCG) mechanisms. 

Besides our theoretical motivation for extending mechanism design to unawareness, we are motivated by practical concerns about the usefulness of explicit mechanisms in reality. For instance, it has been argued that auctions are inappropriate when projects are complex, incompletely designed, and custom-made such as the procurement of new fighter jets, buildings, consulting services, and IT projects. It is claimed that auctions may stifle communication between buyers and sellers, preventing buyers and sellers to use each other's expertise when designing projects (\citet{Goldberg1977, BajariTadelis2001, BajariMcMillanTadelis2008}). The Department of Defense (DoD), the General Service Administration (GSA), and the National Aeronautics and Space Administration (NASA) recently proposed to amend the Federal Acquisition Regulation (FAR) suggesting the prohibition of the use of reverse auctions for complex, specialized, or substantial design and construction services (\citet{OFR2024}).\footnote{When analyzing the use of auctions for procurement at the DoD, \citet{AlperBoning2003} state that auctions work best for well-specified and off-the-shelf commodities but nevertheless remark that the Navy was able to achieve substantial savings with procurement auctions for customized and complex contracts such as CVN camels, i.e., devices for safe separation of ships and piers.} The received view is that for complex projects, negotiations outperform auctions because they can better take advantage of the expertise and know-how of contractors (\citet{Goldberg1977, Sweet1994, BajariMcMillanTadelis2008}). We seek an extension of mechanism design to unawareness to make mechanisms also applicable to complex and incompletely designed projects whose efficient implementation require the pooling of expertise among participants. Our proposed mechanisms combine features of common business practices such as Request for Information (RFI) and Request for Proposals (RFP)\footnote{E.g., Federal Acquisitions Regulation 15.203.} with standard VCG mechanisms such as second price auctions. That is, we combine features of negotiations and traditional mechanism design. While traditionally in economics, negotiations have been interpreted mostly as bargaining over the surplus and thus as a substitute to mechanisms, we view negotiations more as interactively defining the surplus using the expertise of participants, which is complementary to traditional mechanisms implementing and allocating the surplus. 

Extending mechanism design to unawareness has to overcome several obstacles. First, since the mechanism designer herself may be unaware of some payoff type profiles, how could she commit to outcomes for such type profiles? That is, how can she commit to a social choice function without being fully aware of the domain of the social choice function? It is well known that the revelation principle may fail if the mechanism designer cannot commit to the mechanism (e.g., \citet{BesterStrausz2001}).\footnote{In mechanisms under unawareness, there is another potential failure of the revelation principle. Types are always informative about the awareness of an agent because an agent cannot pretend to be a type of which she herself is unaware. In standard mechanism design, type-dependent message sets may lead to a failure of the revelation principle (e.g., \citet{GreenLaffont1986, BullWatson2007, DeneckereSeverinov2008}). Nevertheless, in a separate note, we are able to prove a revelation principle for mechanisms under unawareness.} Yet, the mechanism designer can at least commit to general properties of outcome functions like efficiency given ex-post awareness. Promising to implement an efficient outcome given what transpires in the mechanisms could be verified ex-post by a court of law. That's why we focus on efficient mechanism design under unawareness. We restrict ourselves to the economically relevant class of quasilinear preferences and the notion of utilitarian ex-post efficiency. 

The second obstacle is that, as we will demonstrate in Section~\ref{sec:failure_VCG}, static mechanisms will not suffice for efficient implementation under unawareness. We desire to implement efficiently at the \emph{highest awareness level possible}. The problem is that no agent or the mechanism designer might be aware of everything.=We seek to pool awareness of all agents and aim to implement the social choice that would be utilitarian ex-post efficient at this pooled awareness level. To achieve this, we need dynamic mechanisms that allow prior reported payoff types to be elaborated in light of awareness raised by all agents that is communicated back from the mediator to the agents. 

This leads us to the third obstacle, namely how to model dynamic interactive unawareness. Recent years witnessed laying the foundation for modeling unawareness (\citet{FaginHalpern1988, HMS2006, HMS2013a}). Results by \citet{DekelLipmanRustichini98} imply that standard type spaces preclude unawareness. As remedy, we introduce payoff type spaces that are simplified versions of unawareness type spaces (\citet{HMS2006, HMS2013a}). Payoff types are more than a value or cost. They can be complex descriptions of cost structures and factors affecting valuations involving both verbal, quantitative or pictorial descriptions that might be associated with RFI, RFP, Request for Tender (RFT), and Requests for Quotation (PFQ) in business practice. Important for modelling unawareness is that these descriptions can be ordered by how multi-dimensional, expressive, detailed, fine grained, rich etc. they are. More elaborate descriptions potentially raise awareness of factors that remain hidden or neglected in coarser descriptions of payoff types. The advantage of our abstract formalism is that it provides for a variety of complex descriptions of payoff types while allowing for tractability and generality. Recent years also saw the development of game theory with unawareness (\citet{HMS2013b, MS2014, MS2024, S2021, HalpernRego2014, Feinberg2021}). Our mechanisms introduced in this paper induce dynamic games with unawareness. Dynamic games with unawareness consist of a forest of trees ordered by expressiveness (i.e., the richness of payoff types in our case). An agent's information set at a history in one tree may comprise of histories in a less expressive tree precisely when this agent is unaware of some factors. After nature selected a payoff type and awareness level for each agent, agents report their perceived payoff type in the initial stage of the dynamic direct elaboration mechanism. After that, their awareness may be raised when the mediator provides feedback of the pooled awareness level from first-stage reports of all agents. At this point, agents have the opportunity to elaborate on their previously reported type at the pooled awareness level. This process of elaborations by agents and subsequent communication of pooled awareness by the mediator continues until no agent wants to elaborate any further at which point the mechanism stops and a utilitarian ex-post efficient outcome is implemented. 

Implementation requires a solution concept, a fourth obstacle under unawareness. The dynamics of beliefs and awareness are highly complex, especially in light of awareness changing communication from the mediator. We avoid having to explicitly model the dynamics of beliefs by sticking to the belief-free approach of dominant strategy implementation. However, while dominant strategy implementation may be appropriate for static mechanisms, for dynamic mechanisms it would imply that agents select their entire dominant reporting strategy ex-ante. How can they do this if they are not even aware of all their future information sets and consequently their entire set of strategies? Inspired by conditional dominance for standard games in extensive form by \citet{ShimojiWatson1998} and extended to games with unawareness by \citet{MS2024}, we use implementation in conditional dominant strategies: Conditional on reaching an information set, the agent selects a weak dominant continuation strategy as far as he can anticipate at this information set. A side-benefit of being able to retain the belief-free approach under unawareness is that outcomes are robust to mispecifications of beliefs. 

The last but crucial obstacle is how to incentivize not just the revelation of information but also awareness. Awareness is different from information. An example of payoff type information is ``the cost of producing the item is at least x\$'' while an example of payoff type awareness is ``when digging the foundation near the railway station we may or may not discover a WWII unexploded ordnance whose defusion will cause construction delays''. Note that latter statement involving ``may or may not'' is a tautology. As such, it does not carry information as it does not exclude any event. However, it surely raises awareness of the possibility of such events. In practice, payoff relevant statements often carry both information and awareness like ``our prior experience suggests that when digging near a railways station, we are likely to find a WWII unexploded ordnance whose defusion will create a three day delay of construction at the cost of \$x''. It raises awareness when the buyer has been unaware of the possibility. Moreover, it provides information in stating that such a discovery is ``likely'' and an estimate of the costs. The crucial point though is that when the buyer is unaware, she cannot ask for information about it in the contract tender and agents may not find it in their interest to volunteer such awareness, while when the buyer is aware, she can ask for information (and even infer information from silence; see \citet{Milgrom1981, HMS2021}. Moreover, once the buyer becomes aware of the possibility, she might want to explicitly request information on their cost for such a contingency from \emph{all} agents and thus sharing her newly won awareness among all agents. 

We show that when we pair dynamic direct elaboration mechanisms with transfer schemes inspired by VCG mechanisms (\citet{Groves1973, GrovesLoeb1975}), then we can implement efficiently at the pooled awareness in conditional dominant strategies. It is well known that these transfers incentivize information revelation given any awareness level. However, our transfer schemes also include an additional term that incentivizes raising awareness. This term rewards raising awareness commensurate with its possible effect on social welfare. It does so only if there is a unique agent who raises awareness to the pooled awareness level. Moreover, compared to the standard VCG mechanism without unawareness, an agent has to pay its ``fair'' share of the awareness raising incentives if some other unique agent raises awareness to the final pooled awareness level before she does herself (if such an agent exists). 
 
Next, we investigate properties beyond efficiency. It is well known that standard VCG mechanisms without unawareness do not necessarily satisfy budget balance even without unawareness (\citet{GreenLaffont1979}). We show that awareness pooling does not impose an additional constraint on budget balance. In particular, the characterizing condition for budget balance of our dynamic elaboration VCG mechanisms is the same as for as for standard VCG mechanisms without unawareness (\citet{Holmstroem1977}). The additional incentives for raising awareness are designed to be budget neutral and deviations to lower awareness levels do not impose additional constraints on budget balance. Since budget balance is elusive in the general case (with or without unawareness), we need to look beyond budget balance. Unforeseen contingencies are often used to justify budget overruns. Thus, we are more interested in running no deficit than running a surplus. It is well-known that in standard mechanism design without unawareness, the Clarke or Pivot mechanisms (\citet{Clarke1971}) satisfy no deficit. We show that we can implement a utilitarian ex-post efficient outcome under pooled awareness with no deficit in conditional dominant strategies using a dynamic direct elaboration mechanisms in which transfers are inspired by the Clarke mechanism plus the additional budget neutral term that incentivizes raising awareness. This result is important in the context of unawareness for two reasons: First, as mentioned above, often unforeseen contingencies are used as a justification of cost overruns. Our results imply that ex-ante unforeseen contingencies are not necessarily an excuse for cost overruns. Second, under unawareness the mechanism designer may not be able to anticipate how large subsidies in the mechanism could become. Agents who anticipate this may question whether the mechanism designer could really commit to such large transfers and consequently may not want to report truthfully. By showing that the dynamic elaboration Clarke mechanism satisfies no deficit, we can mitigate this concern.  

Next, we investigate ex-post participation constraints. It is well-known in standard mechanism design that the Clarke mechanism satisfies participation constraints. We show that our dynamic elaboration Clarke mechanism satisfies ex-post participation constraints except that it only holds for ex-post outcomes that are ex-ante anticipated by the agent. While the agent may ex-ante be happy to participate in the mechanism, she may regret it at an interim stage because she comes to realize that she has to pay for the awareness raising incentives. This implies that when the agent is aware of her unawareness (i.e., she realizes that she might be unaware of something without being able to comprehend what she is unaware of; see \citet{S2024}), she might be reluctant to participate in the mechanisms ex-ante because she fears to be penalized when others raise awareness more than she herself does although she has no idea of what other agents could raise awareness of.  

Finally, we focus on the special but important case of procurement under ex-ante unforeseen contingencies. We introduce the dynamic elaboration reverse second price auction that implements an utilitarian ex-post efficient outcome under pooled awareness with budget balance satisfying ex-post participation constraints of all sellers. After the revelation and elaboration stages of the mechanism conclude, the buyer pays the second lowest bid to the seller with the lowest bid who gets the project awarded. The buyer also compensates the unique seller (which may not be the winning seller) who first raises awareness to the pooled awareness level if such a seller exists. 

The paper is organized as follows: In the following section, we introduce payoff type spaces with unawareness and show via a simple example the failure of static VCG mechanisms under unawareness. This is followed by an exposition of dynamic direct elaboration mechanisms in Section~\ref{sec:DDEM} and the proof of efficiency in Section~\ref{sec:DEVCGM}. In Section~\ref{sec:BB}, we characterize budget balance. We also introduce the dynamic elaboration Clarke mechanisms and show no deficit. Participation constraints are discussed in Section~\ref{sec:PC} and procurement under ex-ante unforeseen contingencies is studied in Section~\ref{sec:RSPA}. We conclude in Section~\ref{sec:discussion}.

\subsection{Related Literature\label{sec:literature}} 

Since mechanism design is closely related to contract theory, our paper contributes to the recent literature on contracting under unawareness (e.g., \citet{Lee2008, SommerLoch2009, vonThaddenZhao2012, Auster2013, Filiz-Ozbay2012, GrantKlineQuiggin2012, ChungFortnow2016, AusterPavoni2024, LeiZhao2021, FrancetichSchipper2024}). For instance, \citet{FrancetichSchipper2024} show that screening contracts may not provide sufficient incentives to agents to reveal their awareness. Consequently, the principal may not be able to consider the full set of incentive constraints. In contrast, we show that we can go beyond incomplete contracts and reveal awareness in dynamic direct elaboration mechanisms that provide sufficient incentives for truthful reporting and raising awareness. However, we focus on efficient mechanisms rather than optimal mechanisms, affording us the opportunity of working in the belief-free paradigm of (conditional) dominant strategy implementation and allowing us to conveniently bypass the subtleties of updating beliefs under dynamic awareness. The literature on unawareness in contracting is very different from the earlier literature on indescribable contingencies in contracting. For instance, in \citet{MaskinTirole1999}, agents are fully aware of every particular payoff consequence but cannot ex-ante describe them. In contrast, under unawareness agents may not be fully aware of all payoff consequences in contracting. Closer to mechanism design, \citet{LiSchipper2024} study the seller’s decision to raise bidders’ awareness of characteristics before a second-price auction with entry fees. Optimal entry fees capture an additional unawareness rent due to unaware bidders misperceiving their probability of winning and the price to be paid upon winning. In contrast to our setting, the auctioneer is aware of everything ex-ante. \citet{Piermont2024} studies how a decision maker can incentivize an expert to reveal novel aspects about a decision problem via an iterated revelation mechanism in which at each round the expert decides on whether or not to raise awareness of a contingency that in turn the decision maker considers when proposing a new contract that the expert can accept or reject. This bears some similarity with our dynamic direct elaboration mechanisms. Most importantly, we focus on a multi-agent setting with transferable utilities. \citet{Piermont2024} shows that iterated revelation allows for efficient outcomes to emerge. 

\citet{HerwegSchmidt2020} study a procurement problem with a principal and two agents who may be aware of some design flaws. Our paper differs from theirs in many respects: First, in \citet{HerwegSchmidt2020} agents can raise awareness of realized design flaws, agent's private costs are independent of the design flaw, and fixing the design flaw requires a known common cost that is interim verified by an industry expert. In our model, agents report payoff types, can raise awareness of potential factors affecting payoffs, and individually elaborate how the payoffs change in light of new awareness. Second, \citet{HerwegSchmidt2020} construct an efficient direct mechanism under common awareness and then argue that there is an indirect mechanism that under asymmetric awareness that gives rise to the same outcomes and incentives. This leads them to conclude that there are also corresponding equilibria in these two mechanisms. Yet, an appropriate notion of equilibrium in mechanisms under asymmetric awareness should verify equilibrium behavior w.r.t what outcomes agents anticipate and how behavior is affect by changing anticipations of outcomes during the play. Our approach is more ``direct'': We conduct our entire analysis of efficient conditional dominant strategy implementation in a direct mechanisms under asymmetric awareness thereby modeling every potential deviation from equilibrium behavior from the agent's point view. Finally, \citet{HerwegSchmidt2020} focus on the particular but very relevant application to procurement while we consider more generally the efficient mechanism design problem under asymmetric awareness. 

We study the design of efficient mechanisms in which raising awareness by one agent can via the mediator awareness and thus valuations of other agents. That is, even though we work with private valuations, valuations become endogenously interdependent. We know from \citet{JehielMoldovanu2001} that it is generically impossible to implement ex-post efficient outcome functions in settings with interdependent valuations. Why are we able to nevertheless implement efficiently? The key is that the interdependence in payoffs stems from awareness, which agents can only misreport in one direction. By using transfers that are sufficiently increasing in reported awareness, we incentivize agents to report as much awareness as possible --- namely, their true awareness level. A related result is given by \citet{KraehmerStrausz2024}. We discuss the connection further after the results. 

We make use of unawareness type spaces introduced by \citet{HMS2006, HMS2013a} as well as games with unawareness developed by \citet{HMS2013b, MS2014}, and \citet{S2021}. For alternative approaches, see \citet{FaginHalpern1988, HalpernRego2014, Feinberg2021}, and \citet{BoardChung2021}.

\section{Model\label{sec:model}} 

\subsection{Payoff Types with Unawareness\label{sec:payoff_types}} 

Let $L$ be a finite lattice with order $\trianglerighteq$. Elements of the lattice represent awareness levels. The join of the lattice is denoted by $\bar{\ell} \in L$. Define $L(\ell) := \{\ell' \in L : \ell' \trianglelefteq \ell\}$ for $\ell \in L$. This is the sublattice of $L$ with join $\ell$. The significance of $L(\ell)$ is that an agent with awareness level $\ell$ can only reason about awareness levels in $L(\ell)$.

Fix a nonempty finite set of agents $I$. For each agent $i \in I$, there a collection of nonempty disjoint payoff type spaces $\{T_i^{\ell}\}_{\ell \in L}$. Let $\mathcal{T}_i := \bigcup_{\ell \in L} T_i^{\ell}$. A payoff type $t_i \in \mathcal{T}_i$ is more than just a value for an object. If $t_i \in T_i^{\ell}$, then describing the payoff type also requires at least awareness level $\ell$. We illustrate this feature with the following example.\\

\noindent \textbf{Example 1} Consider the context of procurement. A principal may invite agents to bid on a complex project. Agents do their due diligence and identify relevant items that drive their costs. Suppose there are items $\{a, b, c\}$. Agents may be unaware of some items even after their due diligence. Agent 1 may only be aware of items $\ell_1 = \{a, b\}$ while agent 2 is only aware of items $\ell_2 = \{b, c\}$. In this case, $L$ is isomorphic to the set of all subsets in $2^{\{a, b, c\}}$ and the natural lattice order is induced by set inclusion on $2^{\{a, b, c\}}$. Agent 1 may submit a bid given by the following table $t_1$ while agent 2 may submit a bid given by table $t_2$. 
$$\begin{array}{ccc} t_1 = \begin{array}{|c|c|} \hline \mbox{Item} & \mbox{Cost} \\ \hline
\mbox{a} & 23 \\
\mbox{b} & 41 \\ \hline 
\mbox{Total} & 64 \\ \hline
\end{array} & \quad \quad \quad \quad &
t_2 = \begin{array}{|c|c|} \hline \mbox{Item} & \mbox{Cost} \\ \hline
\mbox{b} & 38 \\
\mbox{c} & 29 \\ \hline 
\mbox{Total} & 67 \\ \hline
\end{array}\end{array}$$ As the example demonstrates, payoff types can be multi-dimensional with varying dimensions. Needless to say, the lattice approach is more general than multi-dimensional payoff type spaces.\hfill $\Box$\\

For any agent $i \in I$ and awareness levels $\ell, k \in L$ with $k \trianglerighteq \ell$, we require a surjective projection $r^{k}_{\ell}: T_i^k \longrightarrow T_i^{\ell}$ such that for all $\ell, \ell', \ell'' \in L$, $\ell'' \trianglerighteq \ell' \trianglerighteq \ell$, $r^{\ell'}_{\ell}(r^{\ell''}_{\ell'}(t_i)) = r^{\ell''}_{\ell}(t_i)$ and $r^{\ell}_{\ell}(t_i) = t_i$. For brevity, we do not index $r^k_{\ell}$ by agents. The projection relates payoff types across awareness levels. Before we illustrate this notion, we also define extensions of payoff types to greater awareness levels. For any agent $i \in I$, awareness level $\ell \in L$, and payoff type $t_i \in T_i^\ell$, we let $t_i^{\uparrow} := \bigcup_{k \trianglerighteq \ell, k \in L} (r_{\ell}^k)^{-1}(t_i)$. That is, $t_i^{\uparrow}$ is the union of inverse images at (weakly) greater awareness levels of payoff type $t_i$. Similarly, for any subset of payoff types in a given payoff type space, we let superscript ``$\uparrow$'' indicate the union of inverse images in payoff type spaces corresponding to greater awareness levels. We illustrate these notions in our prior example.\\ 

\noindent \textbf{Example 1 (Continuation)} Continuing our prior example, consider payoff types $t_1'$ and $t_1''$ of agent 1: 
$$\begin{array}{ccc} t_1' = \begin{array}{|c|c|} \hline \mbox{Item} & \mbox{Cost} \\ \hline
\mbox{a} & 23 \\
\mbox{b} & 41 \\ 
\mbox{c} & 16 \\ \hline 
\mbox{Total} & 80 \\ \hline
\end{array} & \quad \quad \quad \quad &  
t_1'' = \begin{array}{|c|c|} \hline \mbox{Item} & \mbox{Cost} \\ \hline
\mbox{a} & 23 \\
\mbox{b} & 41 \\ 
\mbox{c} & 15 \\ \hline 
\mbox{Total} & 79 \\ \hline
\end{array} \end{array}$$ Both payoff types are described with items in $\{a, b, c\}$. That is, both $t_1', t_1'' \in T_1^{\{a, b, c\}}$. When payoff information on item $c$ is stripped away from payoff type $t_1'$ we obtain payoff type $t_1$. That is, $r^{\{a, b, c\}}_{\{a, b\}}(t_1') = t_1$. Similarly, $r^{\{a, b, c\}}_{\{a, b\}}(t_1'') = t_1$. Thus, $t_1, t_1', t_1'' \in t_1^{\uparrow}$. \hfill $\Box$\\

For any agent $i \in I$, we define $\lambda: \mathcal{T}_i \longrightarrow L$ by $\lambda(t_i) = \ell$ if $t_i \in T_i^{\ell}$. Again, for brevity we do not index $\lambda$ by agents. The function $\lambda$ indicates the awareness level that is required to describe the payoff type. For instance, in Example 1, $\lambda(t_1) = \{a, b\}$ while $\lambda(t_1') = \{a, b, c\}$. 

Our framework is general enough to capture payoff types described by all kinds of formal objects like sets of formulae in a formal language (e.g., \citet{HMS2008}), abstract sets, vectors, matrices, formal concepts, pre-sheaves etc. that may represent verbal, quantitative, or pictorial features of proposals, tenders, quotations, messages etc. in business practice. By abstracting from these particular features, we obtain a theory that works for more than one kind of formalism, ensure tractability, and focus on what is really essential for modeling awareness levels, namely the existence of an order of expressiveness of descriptions.  

For each agent $i \in I$, payoff types and awareness levels are drawn consistently as follows: At awareness level $\bar{\ell}$, nature draws a payoff type $\bar{t}_i \in T_i^{\bar{\ell}}$ in the upmost payoff type space, interpreted as agent $i$'s true payoff type if she were aware of everything, and an awareness level $\ell_i \in L$. Consequently, the agent's perceived payoff type is $r^{\bar{\ell}}_{\ell_i}(t_i)$. That is, agent $i$ can ``miss something'' but he cannot perceive the ``wrong'' payoff type w.r.t. what he is aware. 

More generally, for any awareness level $\ell \in L$, nature draws agent $i$'s corresponding type $r^{\bar{\ell}}_{\ell}(\bar{t}_i)$ and agent $i$'s awareness level $\ell' = \ell_i \wedge \ell$. This means in particular, that if $\ell = \ell_i$, then it is payoff type $r^{\bar{\ell}}_{\ell_i}(t_i)$ and awareness level $\ell_i$ as just discussed. If $\bar{\ell} \trianglerighteq \ell \trianglerighteq \ell_i$, nature draws agent $i$'s corresponding type $r^{\bar{\ell}}_{\ell}(\bar{t}_i)$ and awareness level $\ell_i$. If $\ell \in L$ with $\ell_i \trianglerighteq \ell$, nature draws agent $i$'s corresponding type $r^{\bar{\ell}}_{\ell}(\bar{t}_i)$ and awareness level $\ell$. Finally, if $\ell \in L$ with $\ell \not\trianglerighteq \ell_i$, nature draws agent $i$'s corresponding type $r^{\bar{\ell}}_{\ell}(\bar{t}_i)$ and awareness level $\ell' = \ell_i \wedge \ell$. This specifies perceived payoff types and awareness levels consistently across the type spaces $\{T_i^{\ell}\}_{\ell \in L}$. Notice that if $t_i \in T_i^{\ell}$ for some $\ell \in L$, then the agent's perceived type is always in a type space with (weakly) less awareness than $\ell$. 

\begin{figure}\caption{\label{payoff_type_spaces}}
\begin{center}
\includegraphics[scale=.1]{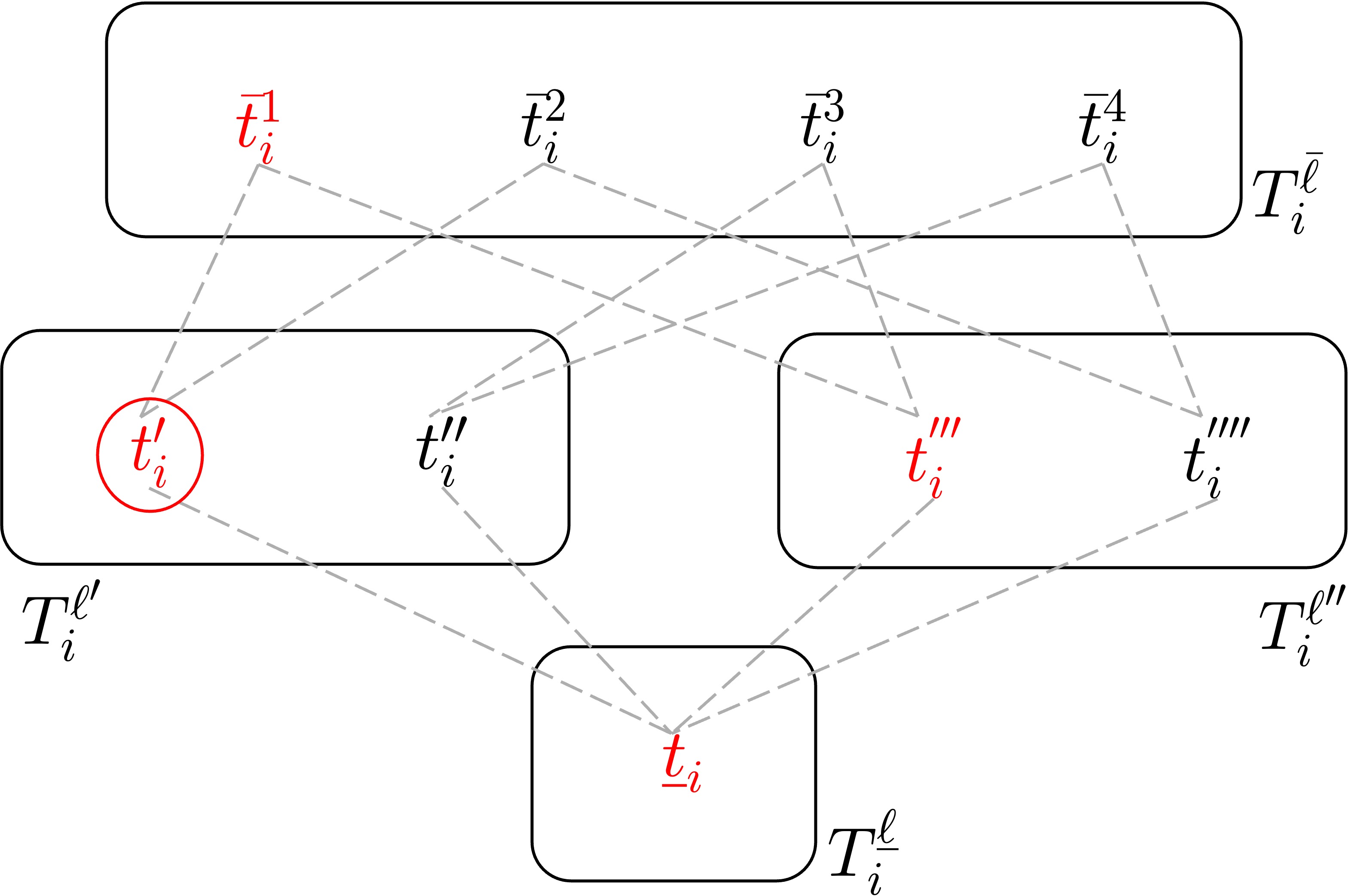}
\end{center}
\end{figure} 

Figure~\ref{payoff_type_spaces} illustrates the payoff type structure. There are four awareness levels $L = \{\bar{\ell}, \ell', \ell'', \underline{\ell}\}$ with $\bar{\ell} \triangleright \ell' \triangleright \underline{\ell}$ and $\bar{\ell} \triangleright \ell'' \triangleright \underline{\ell}$. Associated with each awareness level is a payoff type space for agent $i$. Projections from richer type spaces to poorer type spaces are indicated by dashed lines. Suppose that the payoff type selected by nature in the upmost payoff type space $T_{i}^{\bar{\ell}}$ is $\bar{t}_i^1$ indicated in red. Then the corresponding payoff types selected in the other payoff type spaces follow by projections and are also indicated in red. When the awareness level of agent $i$ selected by nature is $\ell_i = \ell'$, then $t_i'$ is the payoff type perceived by agent $i$. This is indicated with the red circle in type space $T_i^{\ell'}$. 

For any $\ell \in L$, let $\bm{T}^{\ell} := \times_{i \in I} T_i^{\ell}$. Moreover, let $\boldsymbol{\mathcal{T}} := \times_{i \in I} \mathcal{T}_i$. Similarly, we let $\bm{T}^{\ell}_{-i} := \times_{j \in I \setminus \{i\}} T_j^{\ell}$ and $\boldsymbol{\mathcal{T}}_{-i} := \times_{j \in  \in I \setminus \{i\}} \mathcal{T}_j$. 

When agents have the payoff type profile $\bm{t} = (t_i)_{i \in I} \in \boldsymbol{\mathcal{T}}$, their \emph{pooled} awareness level is $\widecheck{\lambda}(\bm{t}) := \bigvee_{i \in I} \lambda(t_i) \in L$, i.e., the join of all agent's awareness levels at payoff type profile $\bm{t}$. Since $L$ is a finite lattice, the join always exists in $L$. Note that $\widecheck{\lambda}(\bm{t})$ may be a greater awareness level than any of the agent's awareness levels. 

For each $\ell \in L$, there is a nonempty compact set of outcomes or allocations $X^{\ell} \subseteq X$. This formulation allows both for awareness of outcomes and awareness that some outcomes are infeasible. 

Each agent $i \in I$ has an upper semi-continuous utility function $u_i: \bigcup_{\ell \in L} X^\ell \times T_i^\ell  \longrightarrow \mathbb{R}$. From this formulation it is clear that we focus on private payoff types. 

To facilitate a commonly used notion of efficiency, we assume that each agent's utility function is quasilinear. I.e., for each $i \in I$, $u_i(x, t_i) := v_i(x_0, t_i) + x_i$ for $x = (x_0, x_1, ..., x_{|I|}) \in X^\ell := X_0^\ell \times \mathbb{R}^{|I|}$ and $t_i \in T_i^\ell$, $\ell \in L$. As usual, $x_0$ describes the physical properties of the outcome while $(x_i)_{i \in I}$ represents the vector of transfers made \emph{to} agents.

We denote the outcome function by $f_0: \boldsymbol{\mathcal{T}} \longrightarrow X_0$. That is, $f_0(\bm{t})$ is the physical outcome prescribed by the outcome function $f_0$ to type profile $\bm{t}$. We require that for any $\bm{t} \in \boldsymbol{\mathcal{T}}$, $f_0(\bm{t}) \in X_0^{\widecheck{\lambda}(\bm{t})}$. That is, the social planner can use joint awareness to select the outcome/allocation. If the reported payoff type profile is $\bm{t}$, then the planner's awareness is $\widecheck{\lambda}(\bm{t})$ and hence outcomes in $X_0^{\widecheck{\lambda}(\bm{t})}$ are selected.\footnote{Typically the planer is not considered as an agent in mechanism design and hence we do not explicitly consider the awareness that the planner may have. Yet, all of our results remain intact when the planner joins her awareness together with the awareness of all agents whenever communicating the pooled awareness level in the dynamic elaboration mechanisms introduced in the next section.} The formulation also makes clear that the social planner cannot necessarily describe the outcome function to agents in advance if the social planner is unaware of something herself. However, we assume that the social planner can commit to abstract properties of outcome functions such as efficiency. 

We generalize utilitarian ex-post efficiency of outcome functions to unawareness by requiring it for every awareness level. 

\begin{definition}\label{def:efficient} The outcome function $f_0$ is \emph{utilitarian ex-post efficient} if for all $\ell \in L$ and $\bm{t} = (t_i)_{i \in I} \in \bm{T}^\ell$,
\begin{align} \sum_{i \in I} v_i(f_0(\bm{t}), t_i) & \geq \sum_{i \in I} v_i(x_0, t_i) \mbox{ for all } x_0 \in X_0^{\ell}.
\end{align}
\end{definition}

\subsection{Failure of Static VCG Mechanisms\label{sec:failure_VCG}} 

In standard mechanism design with quasilinear utilities and private values but without asymmetric unawareness, efficient implementation is almost equivalent to the use of Vickrey-Clark-Groves (VCG) mechanisms. We demonstrate that VCG mechanisms fail to implement efficiently under unawareness, quasilinear utilities, and private values.\\  

\begin{figure}\caption{Payoff Types in Example 2\label{example2}}
\begin{center}
\includegraphics[scale = .1]{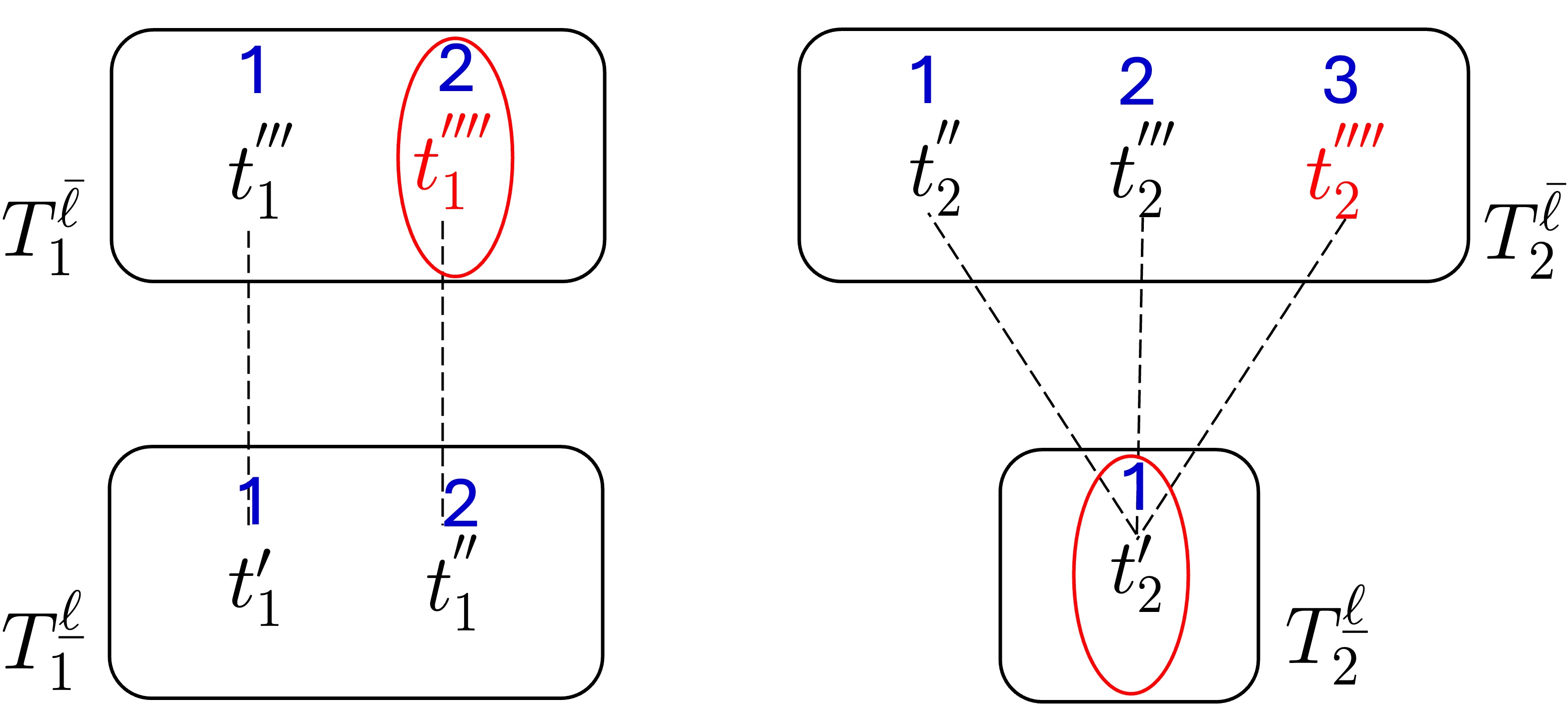}
\end{center}
\end{figure}
\noindent \textbf{Example 2 } Consider two agents, 1 and 2. There are two awareness levels, $\bar{\ell}$ and $\underline{\ell}$ with $\bar{\ell} \triangleright \underline{\ell}$. The payoff type structure is given by Figure~\ref{example2}. For each agent, there are two payoff type spaces, one each corresponding to the two awareness levels. The left payoff types spaces belong to agent 1 while the right ones to agent 2. For each agent's type structure, projections from the larger payoff type space to the smaller one are indicated by dashed lines. For each payoff type, the subscript indices the agent. The superscripts are arbitrary and are made in order to distinguish different payoff types of an agent. 

There is a single object to be allocated. For simplicity, we assume that the value of not getting the object is always zero. The value of getting the object depends on the payoff type. It is written in blue above each payoff type, representing $v_i$. Consider now a payoff type profile $(t_1^{''''}, t_2^{''''})$ and awareness levels $\ell_1 = \bar{\ell}$ and $\ell_2 = \underline{\ell}$ for agents 1 and 2, respectively. In Figure~\ref{example2}, we print the true payoff types in red. Yet, since agents differ in their awareness, we also indicate their corresponding perceived payoff types by red ovals. That is, agent 1 perceives his payoff type to be equal to his actual payoff type $t_1^{''''}$, while agent 2 perceives his payoff type to be $t_2'$ because his awareness level is $\underline{\ell}$. 

Consider running a standard Vickrey second price auction in which the highest bidder wins the object and pays the second highest bid. In the dominant strategy equilibrium of the auction, agents bid truthfully their own perceived payoff type. That is, agent 1 bids 2 and agent 2 bids 1. Consequently, agent 1 obtains the object and pays a price of 1. His (net) utility is 1. Note that awareness is not raised (since agents just submit bids consisting of a value for the object). Clearly, the auction is not efficient since it does not allocate the object to the bidder with the highest true value for the object, which would be bidder 2.

If bidder 1 would have raised awareness of $\bar{\ell}$, then bidder 2 would realize that her value is 3. In such a case, bidder 2 might have bid 3, obtained the object, and paid a price of 2. This leaves bidder 1 with a utility 0, which is worse than his utility of 1 in the Vickrey auction. Thus, in the second price auction, bidder 1 has no incentive to raise bidder 2's awareness.

\subsection{Dynamic Direct Elaboration Mechanisms\label{sec:DDEM}} 

In order to pool awareness among agents and allow agents to provide information on issues they are or became aware, we allow for communication not just \emph{from} agents but also \emph{to} agents. That is, we envision a mediator who receives messages about payoff types from agents, pools the awareness contained in those messages, and sends back messages to agents that potentially raise the awareness of agents to the pooled awareness level. Subsequently, agents may want to elaborate on their prior messages at least to the details of the pooled awareness level. This procedure may be repeated till no agent wants to elaborate further. To this end, we introduce a new class of dynamic mechanisms. 

\begin{definition}[Dynamic Direct Elaboration Mechanism]\label{def:direct_elaboration_mechanism} The \emph{dynamic direct elaboration mechanism} implementing outcome function $f_0$ is defined recursively by the following algorithm:\footnote{Transfers can be arbitrary at this point and will be specified later when we consider particular dynamic direct elaboration mechanisms.}
\begin{itemize}
\item[] Stage $n = 1$: Each agent $i \in I$ must report a type $t_i^1$.

\item[] Stages $n > 1$: Each agent $i$ must report a type $t_i^n \in (t_i^{n-1})^{\uparrow} \cap \left(T_i^{\widecheck{\lambda}(\bm{t}^{n-1})}\right)^{\uparrow}$. 

\item[] Stop: If $t_i^{n+1} = t_i^{n}$ for all $i \in I$, then $f_0(\bm{t}^{n+1})$ is implemented.

\end{itemize}
\end{definition} 

At the first stage, every agent reports a type. At later stages, agents can elaborate on their prior reported types. When agents report the payoff type profile $\bm{t}^{n-1}$ in stage $n - 1$, the mediator awareness level is the pooled awareness level $\widecheck{\lambda}(\bm{t}^{n-1})$. Consequently, he communicates back the pooled awareness level $\widecheck{\lambda}(\bm{t}^{n-1})$ to all agents.\footnote{The current formulation presumes that the mediator, mechanism designer, or social planner has no relevant awareness herself. If she does have relevant awareness, she can incorporate it easily into the message communicated back to agents.} Importantly, all agents receive the same message from the mediator. That is, we can use public messages. At the next stage, each agent $i$ must report a type at least at the pooled awareness level  $T_i^{\widecheck{\lambda}(\bm{t}^{n-1})}$. This report must be consistent with her prior reported type which is implied by the requirement $(t_i^{n-1})^{\uparrow} \cap \left(T_i^{\widecheck{\lambda}(\bm{t}^{n-1})}\right)^{\uparrow}$. For instance, if an agent reported a type whose awareness is strictly below the pooled awareness of all agents' reports, then she must now report a type at least at the pooled awareness level $\widecheck{\lambda}(t_i^{n-1})$ that is consistent with her prior reported type, i.e., a type in $(t_i^{n-1})^{\uparrow}$. However, it could be the case that she has previously pretended to have much less awareness so that the pooled awareness level did not yet incorporate all her awareness. Thus, we allow her to report a type with awareness strictly larger than the pooled awareness level. This motivates the requirement $t_i^n \in (t_i^{n-1})^{\uparrow} \cap \left(T_i^{\widecheck{\lambda}(\bm{t}^{n-1})}\right)^{\uparrow}$ rather than $t_i^n \in (t_i^{n-1})^{\uparrow} \cap T_i^{\widecheck{\lambda}(\bm{t}^{n-1})}$. If an agent did report at the pooled awareness level in the previous stage, then she is not able to change her report in the current stage. The mechanism stops once no agents revise their reports any further. Once it stops, it implements the physical outcome associated to the final reported type profile by $f_0$. \\

\noindent \textbf{Example 1 (Continued)} To continue our Example 1 above, suppose that agent 1 reported in stage 1 payoff type $t_1^1 = t_1$ and agent 2 reported $t_2^1 = t_2$. Pooling awareness leads to $\widecheck{\lambda}(t_1, t_2) = \{a, b, c\}$. At stage 2, both agents must now elaborate their prior reported type and report a type in $T_i^{\{a, b, c\}}$. For instance, agent 1 could report $t_1^2 = t_1'$. Since no further awareness can be revealed after stage 2, the mechanism must conclude after stage 3. \hfill $\Box$\\

Note that awareness and information is transmitted both from agents to the mediator but also from the mediator to the agents. Thus, awareness of agents may change endogenously when interacting in the mechanism. Since agents report directly types and elaborate in later stages one their prior reported types, we call it a ``direct elaboration'' mechanism. Note that by the definition of join of the lattice of spaces, $\widecheck{\lambda}(\bm{t}^{n-1}) \trianglerighteq \lambda(t_i^{n-1})$ for any $i \in I$.  

The mechanism stops when no agent wants to further elaborate on her type. Clearly, since $L$ is finite, the mechanism must stop at some finite stage $n$. Moreover, when it stops at $n$, then $\bm{t}^n \in \bm{T}^{\widecheck{\lambda}(\bm{t}^n)}$. That is, all agents must have reported twice in a row types at the \emph{same} awareness level.  

The dynamic direct mechanism induces a game with unawareness in extensive form \`{a} la \citet{HMS2013b} and \citet{S2021}. Compared to standard games in extensive form, extensive-form games with unawareness feature a forest of game trees with exactly one tree for each level of awareness $\ell \in L$. Moreover, the information set of a player at a history in a tree associated with awareness level $\ell$ may be located in a tree associated with a smaller awareness level $\ell' \trianglelefteq \ell$. For each awareness level $\ell \in L$, in the corresponding tree, nature moves first drawing for each agent $i$ a payoff type in $T_i^{\ell}$ and an awareness level in $L(\ell)$. Across these trees, the draws must be consistent as outlined in the section on payoff types. That is, if payoff type $\bar{t}_i \in T^{\bar{\ell}}$ and awareness level $\ell_i$ are drawn in the tree associated with $\bar{\ell}$, then payoff type $r^{\bar{\ell}}_{\ell}(\bar{t}) \in T^{\ell}$ and awareness level $\ell_i \wedge \ell \in L(\ell)$ are drawn in the tree associated with $\ell$. In any game tree of the forest, the move of nature is followed by histories created by the play of the dynamic direct elaboration mechanism. For any $\ell \in L$, the $\ell$-partial game consists of all game trees associated with awareness levels in $L(\ell)$. The idea is that an agent with awareness level $\ell$ can reason about opponents having awareness associated with any level in $L(\ell)$. At any point during the play, the game perceived by the agent with an awareness level $\ell \in L$ is the $\ell$-partial game. 

To describe information sets, focus first on initial information sets. If nature draws payoff type $\bar{t}_i \in T_i^{\bar{\ell}}$ and awareness level $\ell_i$ for agent $i$, his initial actual information set is $h^{1}_i = r^{\bar{\ell}}_{\ell_i}(\bar{t}_i)$. Here $r^{\bar{\ell}}_{\ell_i}(\bar{t}_i)$ is agent $i$'s initially perceived type. Notice that in this case $h^{1}_i$ is an information set in the tree associated with awareness level $\ell_i$. That is, in games with unawareness, the information set of a player at a history in a tree may be an object of a ``less expressive'' tree that misses aspects of which the agent is unaware. 

In correspondence to the draw of nature in the full game, for any $\ell \in L$ in the $\ell$-partial game, nature draws payoff type $t_i = r^{\bar{\ell}}_{\ell}(\bar{t}_i) \in T_i^{\ell}$ and awareness level $\ell_i \wedge \ell \in L(\ell)$ for agent $i$. The initial information set of agent $i$ in the $\ell$-partial game is $h^{1}_i = r^{\bar{\ell}}_{\ell_i \wedge \ell}(\bar{t}_i) = r^{\ell}_{\ell_i \wedge \ell} (t_i)$. Here $r^{\ell}_{\ell_i \wedge \ell}(t_i)$ is agent $i$'s initially perceived type in the $\ell$-partial game. (Recall that $\ell_i$ is agent $i$'s true initial awareness level selected by nature. It could be that $\ell_i \not\trianglelefteq \ell$. In such a case, the meet notation $\ell_i \wedge \ell$ becomes crucial when specifying the agent's awareness level in the $\ell$-partial game as discussed when introducing the payoff type spaces.) Note that even though $t_i \in T_i^{\ell}$ and thus it is a payoff type in the tree associated with awareness level $\ell$, the corresponding information set and initially perceived payoff type by agent $i$, $h^{1}_i = r^{\ell}_{\ell_i \wedge \ell}(t_i)$, is a payoff type in the tree associated with awareness level $\ell_i \wedge \ell$ which, by definition of the meet, is necessarily weakly less than $\ell$. Again, in games with unawareness the information set associated to a history in one tree may be an object of a less expressive tree. 

Given we have defined all initial information sets, we turn to second stage information sets. For any $i \in I$, fix any initial information set $h_i^1 = t_i$. A successor of $h_i^1$ is defined by $h_i^2 = (t_i', \bm{t}^1)$ with $t_i' \in \left(r_{\lambda(t_i)}^{\lambda(t_i) \vee \widecheck{\lambda}(\bm{t}^1)}\right)^{-1}(t_i)$ and $\bm{t}^1 = (t_i^1, \bm{t}_{-i}^1)$ with $\lambda(t_i^1) \trianglelefteq \lambda(t_i)$. That is, agent $i$ becoming aware in information set $h_i^2$ of $\widecheck{\lambda}(\bm{t}^1)$, joins it with her own awareness $\lambda(t_i)$, and now considers a more elaborate payoff type $t_i' \in \left(r_{\lambda(t_i)}^{\lambda(t_i) \vee \widecheck{\lambda}(\bm{t}^1)}\right)^{-1}(t_i)$. Of course, when communicating at the first stage to the mediator, she could only communicate a payoff type involving awareness not greater than her perceived type at that time $t_i$, i.e., $\lambda(t_i^1) \trianglelefteq \lambda(t_i)$.    

Inductively, for any agent $i \in I$ and $n > 2$, fix a $n-1$ stage information set $h_i^{n-1} = (t_i, (\bm{t}^k)_{k < n-1})$. A successor of $h_i^{n-1}$ is defined by $h_i^n = (t_i', (\bm{t}^k)_{k < n})$ with $t_i' \in \left(r_{\lambda(t_i)}^{\lambda(t_i) \vee \widecheck{\lambda}(\bm{t}^{n-1})}\right)^{-1}(t_i)$, $\bm{t}^{n-1} = (t_i^{n - 1}, \bm{t}_{-i}^{n- 1})$ with $\lambda(t_i^{n-1}) \trianglelefteq \lambda(t_i)$ and $t_j^{n-1} \in (t_j^{n-2})^{\uparrow} \cap (T_j^{\widecheck{\lambda}(\bm{t}^{n - 2})})^{\uparrow}$ for all $j \in I$. The interpretations of the conditions are analogous to the ones for $h_i^2$. The additional condition $t_j^{n-1} \in (t_j^{n-2})^{\uparrow} \cap (T_j^{\widecheck{\lambda}(\bm{t}^{n - 2})})^{\uparrow}$ for all $j \in I$ just says that at the previous stage all agents must have reported a (weak) elaboration of the payoff type reported at the stage before the previous stage since this is implied by the dynamic direct elaboration mechanism. We write $h_i^{n-1} \rightsquigarrow h_i^n$ to indicate that $h_i^n$ is a successor of $h_i^{n-1}$. 

Since $L$ is finite, the mechanism must stop after some finite number of stages. Denote the set of agent $i$'s information sets by $H_i$. 

While we defined each information set essentially as a history, most versions of the dynamic direct elaboration mechanisms we will discuss below will not make use of all this information. Our mechanisms will only make use of the awareness embodied in the reported payoff types profiles but not of the precise payoff type profiles reported along the way except for the last one. Yet, writing information sets with sequences of reported payoff type profiles allows us to keep a uniform notation for all mechanisms we discuss and is notationally simpler than alternatives.\footnote{A more accurate notation for information sets would be $h_i^n = (t_i, (t_i^k, \widecheck{\lambda}(\bm{t}^k))_{k < n})$.}  

We can associate with each information set $h_i \in H_i$ an awareness level. When $h_i^1 = t_i$, agent $i$ has awareness level $\lambda(t_i)$. This motivates an abuse of notation by letting $\lambda(h^1_i) := \lambda(t_i)$ defined by $h_i^1 = t_i$. At later stages $n > 1$, when $h^n_i = (t_i, (\bm{t}^{k})_{k < n})$ occurs, the agent has awareness level $\lambda(t_i)$. (Recall that by the definition of information sets, $t_i$ is already the elaboration of agent $i$'s payoff type in light of awareness raised along the sequence of reported payoff type profiles $(\bm{t}^{k})_{k < n}$.) Again, we abuse notation and write $\lambda(h^n_i) := \lambda(t_i)$. 

For $h_i \in H_i$, let $t_i(h_i) = t_i$ be defined by $h_i^1 = t_i$ if $h_i = h^1_i$ and by $h_i^n = (t_i, (\bm{t}^k)_{k < n})$ if $h_i = h^n_i$. This is the payoff type perceived by agent $i$ in information set $h_i$. 

Since an agent may not necessarily be aware of everything ex-ante, she cannot necessarily envision all of her information sets and choose a strategy that prescribes an action to any of her information sets. At any information set $h_i \in H_i$, agent $i$'s perception of the game is the $\lambda(h_i)$-partial game. For any $\ell \in L$, the set of information sets perceived by her in the $\ell$-partial game is $H_i^{\ell} := \{h_i \in H_i : \lambda(h_i) \trianglelefteq \ell\}$. 

Having specified the representation of information sets in the game with unawareness induced by the dynamic direct elaboration mechanism, we proceed with defining strategies. As standard in game theory, a strategy of an agent assigns to each of her information sets an action. More formally, a (pure) \emph{strategy} of agent $i$ in the game induced by the dynamic direct elaboration mechanism is a mapping $\sigma_i: H_i \longrightarrow \mathcal{T}_i$ such that
\begin{itemize}
\item[(i)] for all $h^1_i \in H_i$, $\sigma_i(h^1_i) \in \bigcup_{\ell \in L(\lambda(h_i))} T_i^{\ell}$,

\item[(ii)] for $n \in \mathbb{N}$ and $h_i^{n} \in H_i$, $\sigma_i(h_i^{n}) \in (t_i^{n-1})^{\uparrow} \cap (T_i^{\lambda(h^n_i)})^{\uparrow}$, where $t_i^{n-1}$ is agent $i$'s previous period's reported payoff type according to $h_i^n$.  
\end{itemize} Property (i) applies to all initial information sets of agent $i$. It says that the agent can report any type that she is aware of at her perceived true type. Property (ii) is a constraint imposed by the dynamic direct elaboration mechanism as it allows only elaborations of the previously reported type.\footnote{This constraint is not necessary for our results. However, it is a natural simplifying assumption on communication, reducing the set of messages over which each agent optimizes and reducing the size of game trees.} An agent can only provide elaborations of her payoff type that she is aware of herself at the information set. She has to elaborate at least at the pooled awareness level $\ell$. Yet, she is also free to elaborate at an even greater awareness level. 

For any awareness level $\ell \in L$ and strategy $\sigma_i$, an $\ell$-partial strategy $\sigma_i^\ell$ is the strategy $\sigma_i$ restricted to information sets in $H_i^\ell$. It assigns an action to each information set of agent $i$ in any game tree of the $\ell$-partial game induced by the mechanism. 

We denote by $\Sigma_i$ the set of agent $i$'s strategies and by $\Sigma_i^{\ell}$ the set of agent $i$'s $\ell$-partial strategies for $\ell \in L$. We denote by $\bm{\Sigma}_{-i} := \times_{j \in I \setminus \{i\}} \Sigma_j$, $\bm{\Sigma}_{-i}^{\ell} := \times_{j \in I \setminus \{i\}} \Sigma_j^\ell$, $\bm{\Sigma} := \times_{i \in I} \Sigma_i$ and $\bm{\Sigma}^{\ell} := \times_{i \in I} \Sigma_i^{\ell}$ the set of strategy profiles of agent $i$'s opponents, the set of $\ell$-partial strategy profiles of agent $i$'s opponents, the set of strategy profiles, and the set of $\ell$-partial strategy profiles, respectively.  

For agent $i$, the \emph{truth-telling and elaboration strategy} $\sigma^*_i$ is defined by: 
\begin{itemize} 
\item For any $h_i^1 \in H_i$, $h^1_i = t_i$ implies $\sigma_i^*(h_i^1) = t_i$.
\item For any $n > 1$, $h_i^n \in H_i$, $h_i^n = (t_i, (\bm{t}^{k})_{k < n})$ implies $\sigma_i^*(h_i^n) = t_i$.  
\end{itemize} 
That is, at each information set, agent $i$ tells her true payoff type as currently perceived at this information set. (Note that at $h^n_i = (t_i, (\bm{t}^{k})_{k < n})$, payoff type $t_i$ incorporates her awareness associated with $\bm{t}^{n-1}$.)

For any agent $i \in I$, awareness level $\ell \in L$, ``initial'' payoff type profile $\bm{t} = (t_j)_{j \in I} \in \bm{T}^{\ell}$, ``initial'' profile of awareness levels $\bm{\ell} = (\ell_j)_{j \in I} \in L(\ell)^{|I|}$, and profile of strategies $\bm{\sigma} = (\sigma_j)_{j \in I} \in \bm{\Sigma}$, we let $H_i(\bm{t}, \bm{\ell}, \bm{\sigma})$ denote the set of agent $i$'s histories reached with $\bm{t}$, $\bm{\ell}$, and $\bm{\sigma}$ (in the $\ell$-partial game). It is defined inductively as follows: 
\begin{eqnarray*} h_i^1 \in H_i(\bm{t}, \bm{\ell}, \bm{\sigma}) & \mbox{if} & h_i^1 = r^{\ell}_{\ell_i}(t_i) \\ 
h_i^2 \in H_i(\bm{t}, \bm{\ell}, \bm{\sigma}) & \mbox{if} & h_i^2 = \left(r^{\ell}_{\ell_i \vee \widecheck{\lambda}((\sigma_j(h_j^1))_{j \in I})}(t_i), (\sigma_j(h^1_j))_{j \in I}\right) \mbox{ for } h^1_j \in H_j(\bm{t}, \bm{\ell}, \bm{\sigma})\\ 
\mbox{ and for } n > 1, & & \\
h_i^n \in H_i(\bm{t}, \bm{\ell}, \bm{\sigma}) & \mbox{ if } & h_i^n = \left(r^{\ell}_{\ell_i \vee \widecheck{\lambda}((\sigma_j(h_j^{n-1}))_{j \in I})}(t_i), (\bm{t}^{k})_{k < n - 1}, (\sigma_j(h_j^{n-1}))_{j \in I}\right) \mbox{ for } \\
& & h_j^{n-1} \in H_j(\bm{t}, \bm{\ell}, \bm{\sigma}) \mbox{ and } h_i^{n-1} \rightsquigarrow h_i^n.
\end{eqnarray*} The initial information set is agent $i$'s perceived payoff type in the $\ell$-partial game. At successive information sets, the initial payoff type is elaborated in light of the pooled awareness raised in the reports induced by the strategy profile at their prior reached information sets. The elaboration is consistent with the initial payoff type profile selected in $\bm{T}^{\ell}$. This feature implies that $H_i(\bm{t}, \bm{\ell}, \bm{\sigma})$ consists of a \emph{unique} path (a unique sequence) of information sets for agent $i$. The requirement $h_i^{n-1} \rightsquigarrow h_i^n$ implies that the sequence  $(\bm{t}^{k})_{k < n - 1}$ in the definition of $h_i^n$ is identical to the sequence of payoff type profiles in $h_i^{n-1}$. 

For any agent $i \in I$, awareness level $\ell \in L$, ``initial'' payoff type profile $\bm{t} = (t_j)_{j \in I} \in \bm{T}^{\ell}$, ``initial'' profile of awareness levels $\bm{\ell} = (\ell_j)_{j \in I} \in L(\ell)^{|I|}$, and strategy $\sigma_i \in \Sigma_i$, we let $H_i(\bm{t}, \bm{\ell}, \sigma_i) := \bigcup_{\bm{\sigma}_{-i} \in \bm{\Sigma}_{-i}} H_i(\bm{t}, \bm{\ell}, \sigma_i, \bm{\sigma}_{-i})$. Similarly, we let $H_i(\sigma_i) := \bigcup_{(\bm{t}, \bm{\ell}) \in \bigcup_{\ell \in L} \bm{T}^{\ell} \times L(\ell)^{|I|}} H_i(\bm{t}, \bm{\ell}, \sigma_i)$. 

For every $\ell \in L$, any ``initial'' payoff type profile $\bm{t} = (t_j)_{j \in I} \in \bm{T}^{\ell}$, ``initial'' profile of awareness levels $\bm{\ell} = (\ell_j)_{j \in I} \in L(\ell)^{|I|}$, and profile of strategies $\bm{\sigma} = (\sigma_j)_{j \in I} \in \bm{\Sigma}$ give rise to a \emph{unique} sequence of payoff type profiles $\tau(\bm{t}, \bm{\ell}, \bm{\sigma}) = (\bm{t}^1, ..., \bm{t}^{n-1}, \bm{t}^n)$ (for some $n > 2$) defined by $\bm{t}^k = (\sigma_j(h_j^k))_{j \in I}$ with $h_j^k \in H_j(\bm{t}, \bm{\ell}, \bm{\sigma})$ for $j \in I$, $k \leq n$, and $\bm{t}^{n-1} = \bm{t}^n$. This sequence is unique because as we observed above $h_j^1$ is unique for each $j \in I$ given $\bm{t}, \bm{\ell}$, and $\bm{\sigma}$, $h_j^2$ is unique for each $j \in I$ given $\bm{t}, \bm{\ell}$, and $\bm{\sigma}$, etc. The requirement $\bm{t}^{n-1} = \bm{t}^n$ means that $\tau(\bm{t}, \bm{\ell}, \bm{\sigma})$ consists of the entire sequence of payoff type profiles until the dynamic direct elaboration mechanisms stops. For the implementation of outcomes, we are especially interested in the last profile of payoff types of any such sequence and denote it by $\tau^*(\bm{t}, \bm{\ell}, \bm{\sigma})$. The set of all sequences is denoted by $\mathfrak{T} := \{\tau(\bm{t}, \bm{\ell}, \bm{\sigma}) : \bm{\sigma} \in \bm{\Sigma}, (\bm{t}, \bm{\ell}) \in \bigcup_{\ell \in L} (\bm{T}^{\ell} \times L(\ell)^{|I|}) \}$ 
 
For any agent $i \in I$ and information set $h_i \in H_i$, define $$\Sigma_i(h_i) := \left\{\sigma_i \in \Sigma_i : \exists (\bm{t}, \bm{\ell}, \bm{\sigma}_{-i}) \in \left(\bigcup_{\ell \in L} (\bm{T}^{\ell} \times L(\ell)^{|I|})\right) \times \bm{\Sigma}_{-i} \  \left(h_i \in H_i(\bm{t}, \bm{\ell}, (\sigma_i, \bm{\sigma}_{-i}))\right)\right\}.$$ This is the set of agent $i$'s strategies that are consistent with information set $h_i$. 

For any agent $i \in I$, let $f_i: \mathfrak{T} \longrightarrow \mathbb{R}$ denote the transfer paid \emph{to} agent $i$ in the dynamic direct elaboration mechanisms. These transfers will depend on the precise versions of the direct elaboration mechanism studied below. In contrast to the outcome function $f_0$ we allow transfers to depend on the entire sequence of reported type profiles. The reason is that it will become important who first reported which awareness level.

We let the social choice function (i.e., the outcome function and transfers) be denoted by $f: \mathfrak{T} \longrightarrow \mathbb{R}$ and defined by $f = (f_0, f_1, ..., f_{|I|})$. That is, for any strategy profile $\bm{\sigma} \in \bm{\Sigma}$, $\ell \in L$, initial profiles of payoff types $\bm{t} \in \bm{T}^{\ell}$, initial profiles of awareness levels $\bm{\ell} \in L(\ell)^{|I|}$, $f(\tau(\bm{t}, \bm{\ell}, \bm{\sigma})) = \left(f_0(\tau^*(\bm{t}, \bm{\ell}, \bm{\sigma})), f_1(\tau(\bm{t}, \bm{\ell}, \bm{\sigma})), ..., f_{|I|}(\tau(\bm{t}, \bm{\ell}, \bm{\sigma}))\right)$.  

\begin{definition}\label{defin:conditional_dominance} The dynamic direct elaboration mechanism truthfully implements the social choice function $f$ in conditional dominant strategies if for all agents $i \in I$, information sets $h_i \in H_i(\sigma^*_i)$, opponents' strategy profiles $\bm{\sigma}_{-i} \in \bm{\Sigma}_{-i}$, initial profiles of payoff types $\bm{t} \in \bm{T}^{\lambda(h_i)}$ (as perceived by $i$ in $h_i$), and initial profiles of awareness levels $\bm{\ell} \in L(\lambda(h_i))^{|I|}$ (as perceived by $i$ in $h_i$) such that $h_i \in H_i(\bm{t}, \bm{\ell}, (\sigma^*_i, \bm{\sigma}_{-i}))$, 
\begin{eqnarray} u_i(f(\tau(\bm{t}, \bm{\ell}, (\sigma_i^*, \bm{\sigma}_{-i})), t_i(h_i))  & \geq & u_i(f(\tau(\bm{t}, \bm{\ell}, (\sigma_i, \bm{\sigma}_{-i})), t_i(h_i)) \label{eqn:conditional_dominance} 
\end{eqnarray} for all $\sigma_i \in \Sigma_i(h_i)$. 
\end{definition}

Conditional dominance strengthens dominance by requiring each agent not only to select a (partial) strategy that is ex-ante dominant but also dominant conditional on each information set. This becomes important when agents cannot anticipate all information sets ex-ante and thus cannot select ex-ante a strategy for the entire game in extensive form. 

\begin{definition}\label{def:pooled_awareness} A outcome function $f_0$ is truthfully implemented at the pooled awareness level if for any $\bm{\bar{t}} \in \bm{T}^{\bar{\ell}}$ and $(\ell_i)_{i \in I} \in L^{|I|}$, $f_0\left(r^{\bar{\ell}}_{\left(\bigvee_{i \in I} \ell_i\right)} (\bm{\bar{t}})\right)$ is implemented. 
\end{definition} 

Recall that $\left(\bm{\bar{t}}, (\ell_i)_{i \in I}\right) \in \bm{T}^{\bar{\ell}} \times L^{|I|}$ represents the move of nature selecting both actual payoff types and awareness levels for all agents. Consequently, $\bigvee_{i \in I} \ell_i$ represents the pooled awareness level.

\section{Efficient Implementation\label{sec:DEVCGM}} 

To implement efficiently at the pooled awareness level, we specify for each agent $i \in I$ the transfers $f_i$ that incentivize both the revelation of information and raising awareness. Revelation of information is achieved via VCG transfers. Raising awareness requires an extra term in the transfer functions. 

\begin{definition}[Dynamic Elaboration VCG Mechanism]\label{def:VCG_mechanism} We say that the dynamic direct elaboration mechanism implementing $f$ is a dynamic elaboration VCG mechanism if $f_0$ is utilitarian ex-post efficient and transfers $f_i$ to agent $i \in I$ are given by for any $(\bm{t}^1, ..., \bm{t}^n) \in \mathfrak{T}$,
\begin{eqnarray}\label{eqn:amended_VCG_transfers1} f_i(\bm{t}^1, ..., \bm{t}^n) := \sum_{j \neq i} v_j(f_0(\bm{t}^n), t_j^n) + y_i^{\widecheck{\lambda}(\bm{t}^n)}(\bm{t}^n_{-i}) + a_i(\bm{t}^1, ..., \bm{t}^n),
\end{eqnarray}
where, for each $\ell$, $y_i^{\ell}: \bm{T}^{\ell}_{-i} \longrightarrow \mathbb{R}$ is an arbitrary function and $a_i(\cdot)$ is defined by:
\begin{eqnarray}\label{eqn:amended_VCG_transfers2}
a_i(\bm{t}^1, ..., \bm{t}^n) :=
\begin{cases}
m_i(\widecheck{\lambda}(\bm{t}^n)) & \mbox{if } i = i^*(\bm{t}^1, ..., \bm{t}^n) \\
-\frac{1}{|I|-1} m_j(\widecheck{\lambda}(\bm{t}^n)) & \mbox{if } j = i^*(\bm{t}^1, ..., \bm{t}^n) \neq i \\
0 & \mbox{otherwise}
\end{cases}
\end{eqnarray}
where 
\begin{eqnarray*} i^*(\bm{t}^1, ..., \bm{t}^n) & := & \left\{i \in I : \exists k \leq n \ \left((\lambda(t_i^k) = \widecheck{\lambda}(\bm{t}^n)) \mbox{ and } \not\exists j \neq i, k' \leq k \ (\lambda(t_j^{k'}) = \widecheck{\lambda}(\bm{t}^n))\right)\right\}
\end{eqnarray*} and $m_i(\ell)$ is defined recursively, as follows: $m_i(\underline{\ell}) :=  0$ and for any $\ell \triangleright \underline{\ell}$,
\begin{eqnarray}\label{eqn:amended_VCG_transfers3} m_i(\ell)  & := &  \max\limits_{\ell' \triangleleft \ell, \bm{t}' \in \bm{T}^{\ell'}, \bm{t} \in \bm{T}^{\ell}}\left\{ \left( m_i(\ell') + v_i(f_0(\bm{t}'), t_i) + \sum_{j\neq i} v_j(f_0(\bm{t}'), t'_j) \right.\right. \nonumber \\ 
& &\left. \left. + y_i^{\ell'}(\bm{t}'_{-i}) - \sum_{j} v_j(f_0(\bm{t}), t_j) - y_i^\ell(
\bm{t}_{-i})\right), 0\right\}.
\end{eqnarray} 
\end{definition}

The mechanism can viewed as a dynamic version of the Vickrey-Clarke-Groves (VCG) mechanisms (\citet{Groves1973, GrovesLoeb1975}). Note that the transfers can be described without necessarily being aware of all payoff type profiles. The mechanism designer commits to implement a utilitarian ex-post efficient outcome given the agents' final reports of payoff types. Each agent is paid the total welfare of others given the final reports of payoff types plus a term that depends on only the opponents' payoff types (and the final pooled awareness level) and additionally a term incentivizing raising of awareness. This last term does not only depend on the final reported payoff type profile but on the sequence of reported payoff type profiles. It matters who reports the pooled awareness level first. Note that $i^*(\bm{t}^1, ..., \bm{t}^n) = \emptyset$ if there is no agent who reports the pooled awareness level (which can happen if the individual awareness levels are incomparable) or if several agents simultaneously report the pooled awareness level first. If $i$ is the unique agent who first reports the pooled awareness level, then $a_i$ receives a payment related to the cost that raising awareness could impose on her utility from the mechanism, net of the $a_i(\cdot)$ term itself. 

\begin{theorem}\label{theo:VCG} The dynamic elaboration VCG mechanism truthfully implements in conditionally dominant strategies a utilitarian ex-post efficient outcome under pooled awareness.  
\end{theorem}

\noindent \textsc{Proof. }  Using equations~(\ref{eqn:amended_VCG_transfers1}), (\ref{eqn:amended_VCG_transfers2}), and~(\ref{eqn:amended_VCG_transfers3}) for transfers, rewrite the defining inequality~(\ref{eqn:conditional_dominance}) of conditional dominant strategy implementation as follows: For all agents $i \in I$, information sets $h_i \in H_i(\sigma_i^*)$, opponents' strategy profiles $\bm{\sigma}_{-i} \in \bm{\Sigma}_{-i}$, initial profiles of payoff types $\bm{t} \in \bm{T}^{\lambda(h_i)}$ (as perceived by $i$ at $h_i$), and initial profiles of awareness levels $\bm{\ell} \in L(\lambda(h_i))^{|I|}$ (as perceived by $i$ at $h_i$) such that $h_i \in H_i(\bm{t}, \bm{\ell}, (\sigma^*_i, \bm{\sigma}_{-i}))$, 
\begin{eqnarray} \lefteqn{v_i(f_0(\tau^*(\bm{t}, \bm{\ell}, (\sigma_i^*, \bm{\sigma}_{-i}))), t_i(h_i)) + \sum_{j \neq i} v_j(f_0(\tau^*(\bm{t}, \bm{\ell}, (\sigma_i^*, \bm{\sigma}_{-i}))), \tau_j^*(\bm{t}, \bm{\ell}, (\sigma_i^*, \bm{\sigma}_{-i})))} \nonumber \\
& & + y_i^{\widecheck{\lambda}(\tau^*(\bm{t}, \bm{\ell}, (\sigma_i^*, \bm{\sigma}_{-i})))}(\tau^*_{-i}(\bm{t}, \bm{\ell}, (\sigma_i^*, \bm{\sigma}_{-i}))) + a_i(\tau(\bm{t}, \bm{\ell}, (\sigma_i^*, \bm{\sigma}_{-i}))) \nonumber \\ 
& \geq & v_i(f_0(\tau^*(\bm{t}, \bm{\ell}, (\sigma_i, \bm{\sigma}_{-i}))), t_i(h_i)) + \sum_{j \neq i} v_j(f_0(\tau^*(\bm{t}, \bm{\ell}, (\sigma_i, \bm{\sigma}_{-i}))), \tau_j^*(\bm{t}, \bm{\ell}, (\sigma_i, \bm{\sigma}_{-i}))) \nonumber \\
&  & + y_i^{\widecheck{\lambda}(\tau^*(\bm{t}, \bm{\ell}, (\sigma_i, \bm{\sigma}_{-i})))}(\tau^*_{-i}(\bm{t}, \bm{\ell}, (\sigma_i, \bm{\sigma}_{-i})))  + a_i(\tau(\bm{t}, \bm{\ell}, (\sigma_i, \bm{\sigma}_{-i})))  \label{eqn:VCGcond}
\end{eqnarray} for all $\sigma_i \in \Sigma_i(h_i)$.

We need to show that this inequality holds in the following cases:
\begin{itemize}
\item[] Case 1: $h_i = h_i^n \in H_i(\sigma_i^*)$ such that $\not\exists j \neq i, k \leq n$ with $\lambda(t_j^{k}) = \lambda(h_i^n)$. That is, at stage $n$, nobody raised awareness to $\lambda(h_i^n)$ in any previous stage and nobody other than $i$ raises it at the current stage $n$ (and $i$ can raise it at the current stage with the truth-telling strategy). 

\item[] Case 2: $h_i = h_i^n \in H_i(\sigma_i^*)$ such that $\not\exists j \neq i, k < n$ with $\lambda(t_j^{k}) = \lambda(h_i^n)$ but $\exists j \neq i$ such that $\lambda(t_j^{n}) = \lambda(h_i^n)$. That is, at stage $n$, nobody raised awareness to $\lambda(h_i^n)$ at any previous stage but there exists agent $j \neq i$ who raises awareness to $\lambda(h_i^n)$ in the current stage as well.  

\item[] Case 3: $h_i = h_i^n \in H_i(\sigma_i^*)$ such that $\exists j \in I$ (possibly $j = i$) and $k < n$ such that $\lambda(t_j^{k}) = \lambda(h_i^n)$. That is, at stage $n$, somebody has raised awareness to $\lambda(h_i)$ already in some earlier stage. 
\end{itemize}

We deal with the cases one-by-one.

\noindent \emph{Case 1: } We have $i = i^*(\tau(\bm{t}, \bm{\ell}, (\sigma_i^*, \bm{\sigma}_{-i})))$ and $a_i(\tau(\bm{t}, \bm{\ell}, (\sigma_i^*, \bm{\sigma}_{-i}))) = m_i(\lambda(h_i))$. 

\noindent \emph{Case 1a - Deviations within the same awareness level: $\sigma_i(h_i) \in T_i^{\lambda(h_i)}$.}  By Definition~\ref{def:direct_elaboration_mechanism}, $\widecheck{\lambda}(\tau^*(\bm{t}, \bm{\ell}, (\sigma_i^*, \bm{\sigma}_{-i}))) = \widecheck{\lambda}(\tau^*(\bm{t}, \bm{\ell}, (\sigma_i, \bm{\sigma}_{-i}))) = \lambda(h_i)$. It follows that in this case $i = i^*(\tau(\bm{t}, \bm{\ell}, (\sigma_i^*, \bm{\sigma}_{-i}))) = i^*(\tau(\bm{t}, \bm{\ell}, (\sigma_i, \bm{\sigma}_{-i})))$ and $a_i(\tau(\bm{t}, \bm{\ell}, (\sigma_i^*, \bm{\sigma}_{-i}))) = a_i(\tau(\bm{t}, \bm{\ell}, (\sigma_i, \bm{\sigma}_{-i})))$. Thus, these terms cancel out at both sides of inequality~(\ref{eqn:VCGcond}). Since the remaining terms represent utilitarian ex-post welfare as anticipated at $h_i$, plus a term that does not depend on $i$'s report within the same awareness level,  inequality~(\ref{eqn:VCGcond}) is now implied by utilitarian ex-post efficiency of the outcome function $f_0$, i.e., inequality~(\ref{def:efficient}). 

\noindent \emph{Case 1b - Deviations to lower awareness levels: $\sigma_i(h_i)\in T_i^{\ell}$ for some $\ell \triangleleft \lambda(h_i)$.} We can rewrite inequality~(\ref{eqn:VCGcond}) as 
\begin{eqnarray} \lefteqn{m_i(\widecheck{\lambda}(\tau^*(\bm{t}, \bm{\ell}, (\sigma_i^*, \bm{\sigma}_{-i})))) - a_i(\tau(\bm{t}, \bm{\ell}, (\sigma_i, \bm{\sigma}_{-i})))} \nonumber \\
& \geq & v_i(f_0(\tau^*(\bm{t}, \bm{\ell}, (\sigma_i, \bm{\sigma}_{-i}))), t_i(h_i)) + \sum_{j \neq i} v_j(f_0(\tau^*(\bm{t}, \bm{\ell}, (\sigma_i, \bm{\sigma}_{-i}))), \tau^*_j(\bm{t}, \bm{\ell}, (\sigma_i, \bm{\sigma}_{-i}))) \nonumber \\ 
& & + y_i^{\widecheck{\lambda}(\tau^*(\bm{t}, \bm{\ell}, (\sigma_i, \bm{\sigma}_{-i})))}(\tau^*_{-i}(\bm{t}, \bm{\ell}, (\sigma_i, \bm{\sigma}_{-i}))) \nonumber \\
& & - v_i(f_0(\tau^*(\bm{t}, \bm{\ell}, (\sigma_i^*, \bm{\sigma}_{-i}))), t_i(h_i)) - \sum_{j \neq i} v_j(f_0(\tau^*(\bm{t}, \bm{\ell}, (\sigma_i^*, \bm{\sigma}_{-i}))), \tau^*_j(\bm{t}, \bm{\ell}, (\sigma_i^*, \bm{\sigma}_{-i}))) \nonumber \\
& & - y_i^{\widecheck{\lambda}(\tau^*(\bm{t}, \bm{\ell}, (\sigma_i^*, \bm{\sigma}_{-i})))}(\tau^*_{-i}(\bm{t}, \bm{\ell}, (\sigma_i^*, \bm{\sigma}_{-i})))
\end{eqnarray} Moreover, the left hand side is greater than or equal to $m_i(\widecheck{\lambda}(\tau^*(\bm{t}, \bm{\ell}, (\sigma_i^*, \bm{\sigma}_{-i})))) - m_i(\widecheck{\lambda}(\tau^*(\bm{t}, \bm{\ell}, (\sigma_i, \bm{\sigma}_{-i}))))$, since either $\sigma_i$ still results in player $i$ being the first to report the final awareness level as anticipated by $i$ at $h_i$ (i.e., $i = i^*(\tau(\bm{t}, \bm{\ell}, (\sigma_i, \bm{\sigma}_{-i})))$), in which case the $a_i(\cdot)$-term is  $m_i(\widecheck{\lambda}(\tau^*(\bm{t}, \bm{\ell}, (\sigma_i, \bm{\sigma}_{-i}))))$, or it does not ($i \neq i^*(\tau(\bm{t}, \bm{\ell}, (\sigma_i, \bm{\sigma}_{-i})))$), in which case the $a_i(\cdot)$-term is non-positive. Note that $m_i(\cdot)$ is non-negative by construction. 
Therefore, it is sufficient to show:
\begin{eqnarray} \lefteqn{\max_{\scriptsize \begin{array}{c} \ell' \triangleleft \widecheck{\lambda}(\tau^*(\bm{t}, \bm{\ell}, (\sigma_i^*, \bm{\sigma}_{-i}))), \\ \bm{t}' \in \bm{T}^{\ell'}, \\ \bm{t}'' \in \bm{T}^{\widecheck{\lambda}(\tau^*(\bm{t}, \bm{\ell}, (\sigma_i^*, \bm{\sigma}_{-i})))}\end{array}\normalsize} \left( m_i(\ell') + v_i(f_0(\bm{t}'), t''_i) + \sum_{j\neq i} v_j(f_0(\bm{t}'), t'_j) + y_i^{\widecheck{\lambda}(\bm{t}')}(\bm{t}'_{-i}) \right.} \nonumber \\ & & - \sum_j v_j(f_0(\bm{t}''), t_j'') - \left. y_i^{\widecheck{\lambda}(\bm{t}'')}(\bm{t}''_{-i}) \right) - m_i(\widecheck{\lambda}(\tau^*(\bm{t}, \bm{\ell}, (\sigma_i^*, \bm{\sigma}_{-i})))) \nonumber \\
& \geq & \max_{\scriptsize \begin{array}{c} \bm{t}' \in \bm{T}^{\widecheck{\lambda}(\tau^*(\bm{t}, \bm{\ell}, (\sigma_i, \bm{\sigma}_{-i})))}, \\ \bm{t}'' \in \bm{T}^{\widecheck{\lambda}(\tau^*(\bm{t}, \bm{\ell}, (\sigma_i^*, \bm{\sigma}_{-i})))}\end{array} \normalsize} \left( v_i(f_0(\bm{t}'), t''_i) + \sum_{j\neq i} v_j(f_0(\bm{t}'), t'_j) + y_i^{\widecheck{\lambda}(\bm{t}')}(\bm{t}'_{-i})\right.\nonumber \\
& & \left. - \sum_j v_j(f_0(\bm{t}''), t_j'') - y_i^{\widecheck{\lambda}(\bm{t}'')}(\bm{t}''_{-i})\right) \nonumber \\
& \geq & v_i(f_0(\tau^*(\bm{t}, \bm{\ell}, (\sigma_i, \bm{\sigma}_{-i}))), t_i(h_i)) + \sum_{j \neq i} v_j(f_0(\tau^*(\bm{t}, \bm{\ell}, (\sigma_i, \bm{\sigma}_{-i}))), \tau^*_j(\bm{t}, \bm{\ell}, (\sigma_i, \bm{\sigma}_{-i}))) \nonumber \\ 
& & + y_i^{\widecheck{\lambda}(\tau^*(\bm{t}, \bm{\ell}, (\sigma_i, \bm{\sigma}_{-i})))}(\tau^*_{-i}(\bm{t}, \bm{\ell}, (\sigma_i, \bm{\sigma}_{-i}))) \nonumber \\
& & - v_i(f_0(\tau^*(\bm{t}, \bm{\ell}, (\sigma_i^*, \bm{\sigma}_{-i}))), t_i(h_i)) - \sum_{j \neq i} v_j(f_0(\tau^*(\bm{t}, \bm{\ell}, (\sigma_i^*, \bm{\sigma}_{-i}))), \tau^*_j(\bm{t}, \bm{\ell}, (\sigma_i^*, \bm{\sigma}_{-i}))) \nonumber \\
& & - y_i^{\widecheck{\lambda}(\tau^*(\bm{t}, \bm{\ell}, (\sigma_i^*, \bm{\sigma}_{-i})))}(\tau^*_{-i}(\bm{t}, \bm{\ell}, (\sigma_i^*, \bm{\sigma}_{-i}))) 
\end{eqnarray} The first inequality follows from the fact that $\widecheck{\lambda}(\tau^*(\bm{t}, \bm{\ell}, (\sigma_i, \bm{\sigma}_{-i})))$ is in the set of awareness levels we maximize over at the l.h.s. of the inequality. The second equality follows from the fact that both $\tau^*(\bm{t}, \bm{\ell}, (\sigma_i^*, \bm{\sigma}_{-i}))$ and $\tau^*(\bm{t}, \bm{\ell}, (\sigma_i, \bm{\sigma}_{-i}))$ are profiles of types we maximize over at the l.h.s. of the inequality. That is, the r.h.s. is just an instance of the expression being maximized. Thus, inequality~(\ref{eqn:VCGcond}) holds by construction of the mechanism in this case. 
 
\noindent \emph{Case 2a - Deviations within the same awareness level: $\sigma_i(h_i) \in T_i^{\lambda(h_i)}$.} In this case, $i^*(\tau(\bm{t}, \bm{\ell}, (\sigma_i, \bm{\sigma}_{-i}))) = i^*(\tau(\bm{t}, \bm{\ell}, (\sigma_i^*, \bm{\sigma}_{-i}))) = \emptyset$. Thus, 
$a_i(\tau(\bm{t}, \bm{\ell}, (\sigma_i, \bm{\sigma}_{-i}))) = \\  a_i(\tau(\bm{t}, \bm{\ell}, (\sigma_i^*, \bm{\sigma}_{-i}))) = 0$.
The same argument as in Case 1a shows that truth-telling is conditionally dominant among deviations within the same awareness level. 

\noindent \emph{Case 2b - Deviations to lower awareness levels: $\sigma_i(h_i)\in T_i^{\ell}$ for some $\ell \triangleleft \lambda(h_i)$.} In this case, there exists $j \neq i$, such that $i^*(\tau(\bm{t}, \bm{\ell}, (\sigma_i, \bm{\sigma}_{-i}))) = j$ while  $i^*(\tau(\bm{t}, \bm{\ell}, (\sigma_i^*, \bm{\sigma}_{-i}))) = \emptyset$. It follows that $a_i(\tau(\bm{t}, \bm{\ell}, (\sigma_i^*, \bm{\sigma}_{-i}))) = 0$ while $a_i(\tau(\bm{t}, \bm{\ell}, (\sigma_i, \bm{\sigma}_{-i}))) \leq 0$. To show inequality~(\ref{eqn:VCGcond}) in this case, it is therefore sufficient to show that
\begin{eqnarray} \lefteqn{v_i(f_0(\tau^*(\bm{t}, \bm{\ell}, (\sigma_i^*, \bm{\sigma}_{-i}))), t_i(h_i)) + \sum_{j \neq i} v_j(f_0(\tau^*(\bm{t}, \bm{\ell}, (\sigma_i^*, \bm{\sigma}_{-i}))), \tau_j^*(\bm{t}, \bm{\ell}, (\sigma_i^*, \bm{\sigma}_{-i})))} \nonumber \\
& & + y_i^{\widecheck{\lambda}(\tau^*(\bm{t}, \bm{\ell}, (\sigma^*_i, \bm{\sigma}_{-i})))}(\tau^*_{-i}(\bm{t}, \bm{\ell}, (\sigma^*_i, \bm{\sigma}_{-i})))  \nonumber \\ 
& \geq & v_i(f_0(\tau^*(\bm{t}, \bm{\ell}, (\sigma_i, \bm{\sigma}_{-i}))), t_i(h_i)) + \sum_{j \neq i} v_j(f_0(\tau^*(\bm{t}, \bm{\ell}, (\sigma_i, \bm{\sigma}_{-i}))), \tau_j^*(\bm{t}, \bm{\ell}, (\sigma_i, \bm{\sigma}_{-i}))) \nonumber \\
&  & + y_i^{\widecheck{\lambda}(\tau^*(\bm{t}, \bm{\ell}, (\sigma_i, \bm{\sigma}_{-i})))}(\tau^*_{-i}(\bm{t}, \bm{\ell}, (\sigma_i, \bm{\sigma}_{-i})))  \label{eqn:reducedinequalityVCG}
\end{eqnarray} 
Since, in this case, some player has already raised awareness $\lambda(h_i) = \widecheck{\lambda}(\tau^*(\bm{t}, \bm{\ell}, (\sigma^*_i, \bm{\sigma}_{-i})))$, we have that $\widecheck{\lambda}(\tau^*(\bm{t}, \bm{\ell}, (\sigma^*_i, \bm{\sigma}_{-i}))) = \widecheck{\lambda}(\tau^*(\bm{t}, \bm{\ell}, (\sigma_i, \bm{\sigma}_{-i})))$. Therefore, inequality~(\ref{eqn:reducedinequalityVCG}) holds by the usual argument that truth-telling is a dominant strategy in the VCG mechanism.

\noindent \emph{Case 3:} In this case, the $a_i(\cdot)$-terms on both the l.h.s. and r.h.s of inequality~(\ref{eqn:VCGcond}) cancel (as they have already been determined, as far as player $i$ is aware). Moreover, player $i$ can only report a type in $T_i^{\lambda(h_i)}$ (types expressing more awareness are infeasible at $h_i$ and types expressing less awareness are ruled out by the requirement that the type announced in stage $n$ must be in $(t_i^{n-1})^{\uparrow} \cap (T_i^{\widecheck{\lambda}(\bm{t}^{n-1})})^{\uparrow}$. That truth-telling is dominant now follows from the standard argument that the VCG mechanism is dominant strategy incentive compatible with constant awareness. 

Finally, with respect to all three cases, recall that deviations to higher or non-comparable awareness levels are not feasible as agent $i$ has no further awareness in $h_i$ than $\lambda(h_i)$. Therefore, we have shown that a truth-telling and elaboration strategy is conditionally dominant in this mechanism.\hfill $\Box$\\

The proof reveals that every agent has an incentive to raise awareness and no agent has an incentive to report a different payoff type given the awareness level. Initially agents report their awareness level and the perceived true payoff type. At a second stage, they elaborate on their previously reported payoff type at the pooled awareness. The final stage just ends the mechanisms. Nothing is revealed in the last stage except that it becomes common knowledge that the mechanism ends. Thus, the dynamic elaboration VCG mechanism can be implemented in three stages. Each agent has even a strict incentive to be the first to raises awareness to the pooled awareness level. 

\begin{proposition} The utilitarian ex-post efficient outcome under pooled awareness is truth-fully implemented in conditional dominant strategies in the game induced by the dynamic elaboration VCG mechanism in at most three stages.
\end{proposition}

\section{Budget Balance\label{sec:BB}}

Ideally, we would like our mechanisms to satisfy further properties beyond utilitarian ex-post efficiency. We are interested in conditions for budget balance. 

\begin{definition}[Budget Balance] We say that the dynamic direct elaboration mechanism with transfer functions $(f_i)_{i \in I}$ is ex-post budget balanced if for all $(\bm{t}^1, ..., \bm{t}^n) \in \mathfrak{T}$, 
\begin{eqnarray} \sum_{i \in I} f_i(\bm{t}^1, ..., \bm{t}^n) & = & 0. \label{eqn:budget_balance}
\end{eqnarray}
\end{definition} 

Note that the $a_i$-terms are budget neutral, i.e., for any $(\bm{t}^1, ..., \bm{t}^n) \in \mathfrak{T}$ we have by construction, $\sum_{j \in I} a_i(\bm{t}^1, ..., \bm{t}^n) = 0$. Thus, we can show that the condition for budget balance of the dynamic elaboration VCG mechanism is identical to the condition for budget balance of the standard VCG mechanism without unawareness (i.e., \citet{Holmstroem1977}) holding for each awareness level. Awareness pooling does not impose an additional constraint on budget balance. 

\begin{proposition}\label{Holmstroem_Theorem_VCG} The dynamic elaboration VCG mechanism implements the social choice function $f$ with budget balance if and only if for every agent $i \in I$ and awareness level $\ell \in L$, there exists a function $g_i^{\ell}: {T}^{\ell}_{-i} \longrightarrow \mathbb{R}$ such that for all $i \in I$ and $\bm{t}^n = (t_i^n, \bm{t}_{-i}^n) \in \bm{T}^\ell$
\begin{align}\label{eqn:BB_VCG} \sum_{i \in I} v_i(f_0(\bm{t}^n), t_i) & = \sum_{i \in I} g_i^{\widecheck{\lambda}(\bm{t}^n)}\left(\bm{t}_{-i}^n\right).
\end{align}
\end{proposition}

\noindent \textsc{Proof. } The proof is an extension of the proof in \citet[Proposition7.10]{Boergers2015}; see also \citet[pp. 53--54]{Milgrom2004}. 

``Only if'': The proof is constructive. Using equations~(\ref{eqn:amended_VCG_transfers1}), (\ref{eqn:amended_VCG_transfers2}), and~(\ref{eqn:amended_VCG_transfers3}) for transfers and~(\ref{eqn:budget_balance}) for budget balance, the dynamic elaboration VCG mechanism is budget balanced if for all $i \in I$ and $(\bm{t}^1, ..., \bm{t}^n) \in \mathfrak{T}$,
\begin{align} \sum_{i \in I} \left(\sum_{j \neq i} v_j(f_0(\bm{t}^n), t_j^n) + y_i^{\widecheck{\lambda}(\bm{t}^n)}(\bm{t}^n_{-i}) + a_i(\bm{t}^1, ..., \bm{t}^n)\right) & = 0.
\end{align} Since the $a_i$-terms are budget neutral by design, this equation is equivalent to 
\begin{align} \sum_{i \in I} \left(\sum_{j \neq i} v_j(f_0(\bm{t}^n), t_j^n) + y_i^{\widecheck{\lambda}(t^n)}(\bm{t}^n_{-i})\right) & = 0
\end{align} which implies that budget balance does not depend on the entire sequence $(\bm{t}^1, ..., \bm{t}^n)$ of reported payoff type profiles but just on the final reported payoff type profile $\bm{t}^n \in \bm{T}^{\widecheck{\lambda}(\bm{t}^n)}$. The last equation is equivalent to
\begin{align*} (|I| - 1) \sum_{i \in I} v_i(f_0(\bm{t}^n), t_i^n) & = - \sum_{i \in I} y_i^{\widecheck{\lambda}(\bm{t}^n)}(\bm{t}^n_{-i}) \\
\sum_{i \in I} v_i(f_0(\bm{t}^n), t^n_i) & = - \sum_{i \in I} \frac{y_i^{\widecheck{\lambda}(\bm{t}^n)}(\bm{t}^n_{-i})}{|I| - 1}.
\end{align*} Set
\begin{align*} g_i^{\widecheck{\lambda}(\bm{t}^n)}\left(\bm{t}_{-i}^n\right) & := - \frac{y_i^{\widecheck{\lambda}(\bm{t}^n)}(\bm{t}^n_{-i})}{|I| - 1},
\end{align*} obtaining equation~(\ref{eqn:BB_VCG}). This shows necessity.

``If'': Assume equation~(\ref{eqn:BB_VCG}) and define
\begin{align} y_i^{\widecheck{\lambda}(\bm{t}^n)}(\bm{t}^n_{-i}) & := - (|I| - 1) g_i^{\widecheck{\lambda}(\bm{t}^n)}\left(\bm{t}_{-i}^n\right). \label{eqn:defin_g}
\end{align} Then
\begin{align*} \lefteqn{\sum_{i \in I} \left(\sum_{j \neq i} v_j(f_0(\bm{t}), t_j) - (|I| - 1) g_i^{\widecheck{\lambda}(\bm{t}^n)}\left(\bm{t}_{-i}^n\right) + a_i(\bm{t}^1, ..., \bm{t}^n)\right)} \\
& = \sum_{i \in I} \left(\sum_{j \neq i} v_j(f_0(\bm{t}), t_j) - (|I| - 1) g_i^{\widecheck{\lambda}(\bm{t}^n)}\left(\bm{t}_{-i}^n\right) \right) \\
& = (|I| - 1) \sum_{i \in I} v_i(f_0(\bm{t}), t_i) - (|I| - 1) \sum_{i \in I} g_i^{\widecheck{\lambda}(\bm{t}^n)}\left(\bm{t}_{-i}^n\right) = 0
\end{align*} where the second line follows from budget neutrality of $a_i$-terms and the last line follows from equation~(\ref{eqn:BB_VCG}). \hfill $\Box$\\

While the proposition demonstrates that the incentives for awareness pooling do not necessarily add additional constraints on budget balance, it does not mean that budget balance is easy to satisfy with dynamic elaboration VCG mechanisms.

\subsection{No Deficit\label{sec:ND}} 

In classical mechanism design, it is well known that the VCG mechanisms like the Groves mechanisms can not satisfy budget balance in general (\citet{GreenLaffont1979}). We may want to look for weaker requirements. Not being budget balanced means that the mechanisms can run a deficit or surplus. In the context of unawareness, we are interested more in running no deficit than running no surplus for two reasons: First, if the mechanisms runs a deficit larger than the mechanism designer anticipated, agents may suspect that the mechanism designer is not committed to the mechanisms and become reluctant to truth-fully report their payoff types. Second, unforeseen contingencies are often used to justify budget overruns. But are deficits really inevitable in the presence of unawareness? 

Recall that we used the convention that transfers $f_i$ denote transfers \emph{to} agent $i$. The following property is sometimes also called weak budget balance (e.g., \citet{ShohamLeyton-Brown2012}). 

\begin{definition}[No deficit]\label{def:no_deficit} We say that the dynamic direct elaboration mechanism with transfer functions $(f_i)_{i \in I}$ satisfies no deficit if for all $(\bm{t}^1, ..., \bm{t}^n) \in \mathfrak{T}$, 
\begin{eqnarray} \sum_{i \in I} f_i(\bm{t}^1, ..., \bm{t}^n) & \leq & 0. \label{eqn:no_transfers}
\end{eqnarray}
\end{definition} 

In classical mechanism design, it is well known that a Groves mechanism may run a deficit, but the Clarke (or Pivotal) mechanism, in which each agent pays the negative externality that they impose on other agents through their effect on the social choice, does not. Therefore, the idea is to design a version of dynamic elaboration VCG mechanisms with Clarke transfers and show that it does not run a deficit. 

To define the mechanism, we have to specify externalities. For any agent $i \in I$, define the $(-i)$-utilitarian ex-post efficient outcome function $f^{-i}_0: \boldsymbol{\mathcal{T}} \longrightarrow X_{0}$ by $f^{-i}_0(\bm{t}) = \arg \max_{x_{0} \in X^{\widecheck{\lambda}(\bm{t})}_{0}} \sum_{j \neq i} v_j(x_{0}, t_j)$. That is, for each profile of payoff types $\bm{t} \in \boldsymbol{\mathcal{T}}$, $f^{-i}_0$ maximizes utilitarian ex-post welfare taking into account only the value functions of agent $i$'s opponents. Note that different from restricted outcome functions in standard Clarke mechanism the argument of $f^{-i}_0$ is the \emph{full} profile of payoff types $\bm{t} = (t_j)_{j \in I}$ rather than just $\bm{t}_{-i}$. The reason is that in order to compute $f^{-i}_0$ we need the awareness of \emph{all} agents, not just agents $j \neq i$: Although agent $i$ value function is not considered when evaluating the social welfare of an outcome, for agents $j \neq i$, the efficiency of this outcome is still evaluated at the pooled awarness level.

\begin{definition}[Dynamic Elaboration Clarke Mechanism]\label{def:clarke_mechanism} We say that the dynamic direct elaboration mechanism implementing $f$ is a dynamic elaboration Clarke mechanism if it is a dynamic elaboration VCG mechanisms with, for all $\bm{t}^n \in \bm{T}^{\widecheck{\lambda}(\bm{t}^n)}$ and $i \in I$, 
$$y_i^{\widecheck{\lambda}(\bm{t}^n)}(\bm{t}^n_{-i}) := - \sum_{j \neq i} v_j(f^{-i}_0(\bm{t}^n), t^n_j).$$
\end{definition}

Observe that for any $i \in I$, if $\sum_{j \neq i} v_j(f^{-i}_0(\bm{t}^n), t^n_j) \geq \sum_{j \neq i} v_j(f_0(\bm{t}^n), t_j^n)$ then $f_i(\bm{t}^1, ..., \bm{t}^n) - a_i(\bm{t}^1, ..., \bm{t}^n)$ is non-positive. Since $f^{-i}_0(\bm{t}^n)$ maximizes the sum of values over $j \in I \setminus \{i\}$, we have $\sum_{j \neq i} v_j(f^{-i}_0(\bm{t}^n), t^n_j) \geq \sum_{j \neq i} v_j(f_0(\bm{t}^n), t_j^n)$. Together with budget neutrality of the $a_i$-terms, this implies now that the mechanism satisfies no deficit. Utilitarian ex-post efficiency is implied by Theorem~\ref{theo:VCG}. 

\begin{theorem}\label{theo:Clarke} The dynamic elaboration Clarke mechanism truthfully implements in conditionally dominant strategies a utilitarian ex-post efficient outcome under pooled awareness with no deficit.  
\end{theorem}

We illustrate the dynamic elaboration Clarke mechanisms with two examples. In the first example, neither agent will announce the pooled awareness level as the pooled awareness level is the join that is strictly greater than any agent's awareness.\\ 

\noindent \textbf{Example 1 (Continuation)} Consider again Example 1. In the conditional dominant solution to the dynamic elaboration Clarke mechanism, agents report in the first stage respectively, 
$$\begin{array}{ccc} t_1 = \begin{array}{|c|c|} \hline \mbox{Item} & \mbox{Cost} \\ \hline
\mbox{a} & 23 \\
\mbox{b} & 41 \\ \hline 
\mbox{Total} & 64 \\ \hline
\end{array} & \quad \quad \quad \quad & t_2 = \begin{array}{|c|c|} \hline \mbox{Item} & \mbox{Cost} \\ \hline
\mbox{b} & 38 \\
\mbox{c} & 29 \\ \hline 
\mbox{Total} & 67 \\ \hline
\end{array}
\end{array}$$ 
Since $\lambda(t_1) = \{a, b\}$ and $\lambda(t_2)=\{b, c\}$, agents are made aware of the join $\widecheck{\lambda}(t_1, t_2) = \{a, b, c\}$ and are invited to report an elaborated type in $T_i^{\{a, b, c\}}$, $i \in {1, 2}$, respectively. Extending the example slightly, let agents report truthfully their elaborations, respectively, 
$$\begin{array}{ccc} t'_1 = \begin{array}{|c|c|} \hline \mbox{Item} & \mbox{Cost} \\ \hline
\mbox{a} & 23 \\
\mbox{b} & 41 \\ 
\mbox{c} & 16 \\ \hline 
\mbox{Total} & 80 \\ \hline
\end{array} & \quad \quad \quad \quad & t'_2 = \begin{array}{|c|c|} \hline \mbox{Item} & \mbox{Cost} \\ \hline
\mbox{a} & 19 \\
\mbox{b} & 38 \\
\mbox{c} & 29 \\ \hline 
\mbox{Total} & 86 \\ \hline
\end{array}
\end{array}$$ 
To complete the example, suppose that the possible physical outcomes (at any awareness level) are that the good is produced by agent 1, the good is produced by agent 2 or the good is not produced, i.e., $X_0 = \{1, 2, \emptyset\}$. Finally, let there be a third agent, a buyer, who always values the good at 100 no matter whether it comes from agent 1 or agent 2, i.e., $v_3(1, t_3) = v_3(2, t_3) = 100$ and $v_3(\emptyset, t_3) = 0$ for any $t_3$. 

Then, the efficient decision at the updated type profile is for the good to be produced by agent 1 (at a total cost of 80). Since no player is the first to announce the joint awareness level, the $a_i(\cdot)$-term is $0$ for each agent. Agent 3 is pivotal in the sense that if agent 3's valuation is not considered, the good would not be produced at all. We have for all $\bm{t}$, $f(\bm{t}) = 1$, $f^{-1}(\bm{t}) = f^{-2}(\bm{t}) = 1$, and $f^{-3}(\bm{t}) = \emptyset$. The transfers to agent 1 are $100 - 0 - (100 - 0) + 0 = 0$, to agent 2 they are $20 - (100 - 80) - 0 = 0$, and to agent 3 are $-80 - 0 + 0 = -80$. Thus, the mechanism runs a surplus of $80$.\footnote{Note that the mechanism does not satisfy agent 1's ex-post participation constraints. This is true of the Clarke mechanism in this context even without any unawareness. It is known that the Clarke mechanisms may not necessarily satisfy ex-post participation constraints. We will analyze participation constraints in the next section.} \hfill $\Box$\\

Note that in this example awareness did not improve perceived utilitarian ex-post welfare. Nevertheless, we are able to raise awareness to the pooled awareness level with out mechanisms. 

The next example illustrates a case in which the novel $a_i$-terms are non-trivial. \\

\noindent \textbf{Example 2 (Continuation)} Consider again Example 2, as depicted in Figure~\ref{example2}. The dynamic Clarke mechanism would proceed as follows: In the first stage, agent 1 announces $t'''_1$ (with $\lambda(t'''_1) = \bar{\ell}$) and agent 2 announces $t'_2$ (with $\lambda(t'_2) = \underline{\ell}$). In the second stage, agent 1 repeats his announcement and agent 2 reports $t''''_2$ (with $\lambda(t''''_2) = \bar{\ell})$, his elaboration of $t_2'$. At the third stage both agents repeat the previous announcement and the mechanisms stops. Agent 2 receives the good. Since agent 1 was the unique first agent to announce the joint awareness, agent 2 pays $m_1(\bar{\ell})$ to agent 1 (in addition to making the standard Clarke payment to the mechanism, latter amounting to the payment of $2$ like in the second price auction in this example). For computing $m_1(\bar{\ell})$, assume for simplicity that the good is always given to agent 1 when both agents have the same value for the good. Then 
\begin{eqnarray*} m_1(\bar{\ell}) & = & \max_{\underline{\bm{t}} \in T^{\underline{\ell}}, \overline{\bm{t}} \in T^{\bar{\ell}}} \left(v_1(f_0(\underline{\bm{t}}), \underline{t}_1) + v_2(f_0(\underline{\bm{t}}), \underline{t}_2) - v_2(f^{-1}_0(\underline{t}_2), \underline{t}_2) \right.\\
& & \left. - \left(v_1(f_0(\overline{\bm{t}}), \overline{t}_1) + v_2(f_0(\overline{\bm{t}}), \overline{t}_2) - v_2(f^{-1}_0(\overline{t}_2), \overline{t}_2)\right)\right)
\end{eqnarray*} 
The term in the second line is non-negative. In the maximum, it is equal to zero. This is the case when we take $\overline{t}_2 = t'''_2$ and $\overline{t}_1 = t_1''''$. In such a case, the term is $2 + 0 - 2 = 0$. The r.h. term in the first line is at most $1$, which is achieved when $\underline{\bm{t}} = (t''_1, t'_2)$ with associated values equal to $(2, 1)$, and noting that the only other type profile at this awareness level makes the term zero. Thus, $m_1(\bar{\ell}) = 1$. Note that this is non-negative, so the restriction that $m_i(\cdot)$ must be non-negative is not binding. 

The additional terms in the transfers are just the payments from a second price auction at the joint awareness level (0 from agent 1 and 2 from agent 2). Therefore, agent 1 receives 1 and agent 2 pays 2. The mechanism runs a surplus of 1. Agent 1 has no incentive to conceal her awareness, since she gets a utility of 1 no matter whether she reveals it or not. \hfill $\Box$\\

\section{Participation Constraints\label{sec:PC}}

In this section, we study voluntary participation in dynamic elaboration Clarke mechanisms. We assume that agents have an outside option yielding them a payoff of zero for sure. As for VCG mechanisms without unawareness, we focus on ex-post participation constraints. In the presence of unawareness, at the onset of the mechanism, when they might have to sign up for participation, agents do not necessary anticipate all ex-post outcomes. That's why we will distinguish between satisfying ex-post participation constraints and satisfying just \emph{ex-ante anticipated} ex-post participation constraints. 

\begin{definition}[ex-post participation constraints] 
A dynamic direct elaboration mechanism implementing the social choice function $f$ satisfies ex-post participation constraints if for all agents $i \in I$, information sets $h_i \in H_i(\bm{\sigma}^*)$, initial profiles of payoff types $\bm{t} \in \bm{T}^{\lambda(h_i)}$ (as perceived by $i$ at $h_i$), and initial profiles of awareness levels $\bm{\ell} \in L(\lambda(h_i))^{|I|}$ (as perceived by $i$ at $h_i$) such that $h_i \in H_i(\bm{t}, \bm{\ell}, \bm{\sigma}^*)$, 
$$u_i(f(\tau(\bm{t}, \bm{\ell}, \bm{\sigma}^*)), t_i(h_i)) \geq 0.$$ 
\end{definition} 

\begin{definition}[ex-ante anticipated ex-post participation constraints]
The dynamic direct elaboration mechanism satisfies ex-ante anticipated ex-post participation constraints if for every agent $i \in I$ it satisfies ex-post participation constraints for all initial information sets $h_i^1 \in H_i(\bm{\sigma}^*)$.
\end{definition} 

In standard mechanism design with unawareness, participation constraints are typically not satisfied by the Clarke mechanisms without further assumptions. One such a assumption in the literature are non-negative valuations:

\begin{assumption}[Non-negative valuations]\label{ass:no_negative_externalities} For any $\ell \in L$, $\bm{t} = (t_i)_{i \in I} \in \bm{T}^{\ell}$, $i \in I$, and $x_0 \in X_{0}^{\ell}$, $v_i(x_0, t_i) \geq 0$. 
\end{assumption}

This assumption would be satisfied for instance in the context of the sale of a good via an auction or in the context of public goods/projects. 

\begin{proposition}\label{prop: pc} Assumption~\ref{ass:no_negative_externalities} implies that the dynamic elaboration Clarke mechanism satisfies ex-ante anticipated ex-post participation constraints.
\end{proposition}

\noindent \textsc{Proof. }  In the conditionally dominant strategy outcome, for all agents $i \in I$, initial information sets $h_i^1 \in H_i(\bm{\sigma}^*)$, initial profiles of payoff types $\bm{t} \in \bm{T}^{\lambda(h^1_i)}$ (as perceived by $i$ at $h^1_i$), and initial profiles of awareness levels $\bm{\ell} \in L(\lambda(h^1_i))^{|I|}$ (as perceived by $i$ at $h^1_i$) such that $h^1_i \in H_i(\bm{t}, \bm{\ell}, \bm{\sigma}^*)$,
\begin{eqnarray*} \lefteqn{u_i(f(\tau(\bm{t}, \bm{\ell}, \bm{\sigma}^*), t_i(h^1_i))} \\  
& = & v_i(f_0(\tau^*(\bm{t}, \bm{\ell}, \bm{\sigma}^*)), t_i(h_i^1)) + f_i(\tau(\bm{t}, \bm{\ell}, \bm{\sigma}^*)) \\
& = & v_i(f_0(\tau^*(\bm{t}, \bm{\ell}, \bm{\sigma}^*)), t_i(h_i^1)) +  \sum_{j \neq i} v_j(f_0(\tau^*(\bm{t}, \bm{\ell}, \bm{\sigma}^*)), \tau^*_j(\bm{t}, \bm{\ell}, \bm{\sigma}^*)) \\
& & - \sum_{j \neq i} v_j(f^{-i}_0(\tau^*(\bm{t}, \bm{\ell}, \bm{\sigma}^*)), \tau^*_j(\bm{t}, \bm{\ell}, \bm{\sigma}^*)) + a_i(\tau(\bm{t}, \bm{\ell}, \bm{\sigma}^*)) \\
& = & \sum_{j \in I} v_j(f_0(\tau^*(\bm{t}, \bm{\ell}, \bm{\sigma}^*)), \tau^*_j(\bm{t}, \bm{\ell}, \bm{\sigma}^*)) - \sum_{j \neq i} v_j(f^{-i}_0(\tau^*(\bm{t}, \bm{\ell}, \bm{\sigma}^*)), \tau^*_j(\bm{t}, \bm{\ell}, \bm{\sigma}^*)) + a_i(\tau(\bm{t}, \bm{\ell}, \bm{\sigma}^*)) \\
& \geq & \sum_{j \in I} v_j(f_0(\tau^*(\bm{t}, \bm{\ell}, \bm{\sigma}^*)),\tau^*_j(\bm{t}, \bm{\ell}, \bm{\sigma}^*)) - \sum_{j \neq i} v_j(f^{-i}_0(\tau^*(\bm{t}, \bm{\ell}, \bm{\sigma}^*)), \tau^*_j(\bm{t}, \bm{\ell}, \bm{\sigma}^*))
\end{eqnarray*} where equality follows from $t_i(h_i^1) = \tau^*_i(\bm{t}, \bm{\ell}, \bm{\sigma}^*)$ and the last inequality follows from $\lambda(h^1_i) = \widecheck{\lambda}(\tau^*(\bm{t}, \bm{\ell}, \bm{\sigma}^*))$ implying $a_i(\tau(\bm{t}, \bm{\ell}, \bm{\sigma}^*)) \geq 0$. (Recall that if $\bm{t} \in \bm{T}^{\lambda(h^1_i)}$ and $\bm{\ell} \in L(\lambda(h^1_i))^{|I|}$, then $\tau^*(\bm{t}, \bm{\ell}, \bm{\sigma}^*) \in \bm{T}^{\lambda(h^1_i)}$. If $i = i^*(\tau(\bm{t}, \bm{\ell}, \bm{\sigma}^*))$, then $a_i(\tau(\bm{t}, \bm{\ell}, \bm{\sigma}^*)) \geq 0$. Otherwise, if $i^*(\tau(\bm{t}, \bm{\ell}, \bm{\sigma}^*)) = \emptyset$ (because several players raise the pooled awareness level at the first stage), then $a_i(\tau(\bm{t}, \bm{\ell}, \bm{\sigma}^*)) = 0$. The case $k = i^*(\tau(\bm{t}, \bm{\ell}, \bm{\sigma}^*)) \neq i$ is ruled out by playing $\sigma_i^*$.) That is, agent $i$ at $h_i^1$ possesses already the anticipated pooled awareness level and raises it in the truth-telling solution. 

Note $f_0(\tau^*(\bm{t}, \bm{\ell}, \bm{\sigma}^*))$ could pick $f^{-i}_0(\tau^*(\bm{t}, \bm{\ell}, \bm{\sigma}^*))$ if latter is the utilitarian ex-post welfare maximizing outcome when considering the utility of all agents in $I$. Thus, 
\begin{eqnarray*} \sum_{j \in I} v_j(f_0(\tau^*(\bm{t}, \bm{\ell}, \bm{\sigma}^*)), \tau^*_j(\bm{t}, \bm{\ell}, \bm{\sigma}^*)) & \geq & \sum_{j \in I} v_j(f^{-i}_0(\tau^*(\bm{t}, \bm{\ell}, \bm{\sigma}^*)), \tau^*_j(\bm{t}, \bm{\ell}, \bm{\sigma}^*))
\end{eqnarray*} Moreover, by Assumption~\ref{ass:no_negative_externalities}, 
\begin{eqnarray*} \lefteqn{\sum_{j \neq i} v_j(f^{-i}_0(\tau^*(\bm{t}, \bm{\ell}, \bm{\sigma}^*)), \tau^*_j(\bm{t}, \bm{\ell}, \bm{\sigma}^*)) + v_i(f^{-i}_0(\tau^*(\bm{t}, \bm{\ell}, \bm{\sigma}^*)), \tau^*_i(\bm{t}, \bm{\ell}, \bm{\sigma}^*))} \\ & & \quad \quad \quad \quad \quad \quad \quad \quad \quad \quad \quad \quad \geq \sum_{j \neq i} v_j(f^{-i}_0(\tau^*(\bm{t}, \bm{\ell}, \bm{\sigma}^*)), \tau^*_j(\bm{t}, \bm{\ell}, \bm{\sigma}^*))
\end{eqnarray*} Thus, 
\begin{eqnarray*} \sum_{j \in I} v_j(f_0(\tau^*(\bm{t}, \bm{\ell}, \bm{\sigma}^*)), \tau^*_j(\bm{t}, \bm{\ell}, \bm{\sigma}^*)) & \geq & \sum_{j \neq i} v_j(f^{-i}_0(\tau^*(\bm{t}, \bm{\ell}, \bm{\sigma}^*)), \tau^*_j(\bm{t}, \bm{\ell}, \bm{\sigma}^*))
\end{eqnarray*} which together with the arguments above implies 
\begin{eqnarray*} u_i(f(\tau(\bm{t}, \bm{\ell}, \bm{\sigma}^*)), t_i(h^1_i)) & \geq & 0
\end{eqnarray*} \hfill $\Box$ \\

While the dynamic elaboration Clarke mechanisms satisfies \emph{ex-ante anticipate} ex-post participation constraints under Assumption~\ref{ass:no_negative_externalities}, it does not satisfy ex-post participation constraints in general because during the play of the mechanisms agent $i$ may become aware due to some agent raising awareness to a level strictly larger than $i$'s awareness level. This is illustrated in the following example.\\

\begin{figure}\caption{Payoff Types in Example 3\label{example4}}
\begin{center}
\includegraphics[scale = .1]{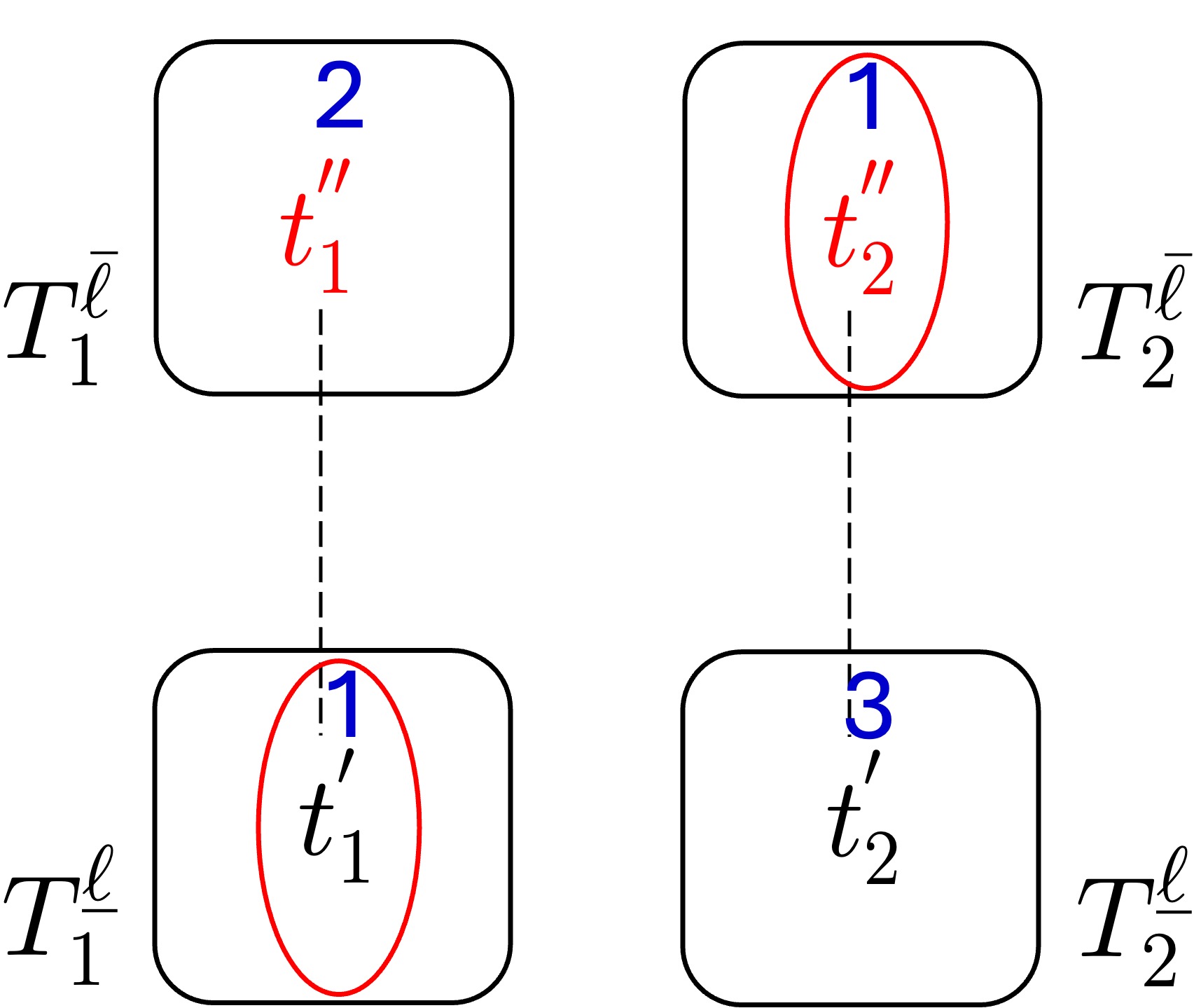}
\end{center}
\end{figure}
\noindent \textbf{Example 4} Similar to Example 3, there are two agents and two awareness levels $L = \{\underline{\ell}, \bar{\ell}\}$. For each agent and awareness level, the payoff type space is a singleton. Again, this will make the computation of the maximum in the definition of $m_i$ for $i \in \{1, 2\}$ trivial. The payoff type spaces are given in Figure~\ref{example4}. There is a single object to be allocated. When an agent obtains the object, the valuation for the object is printed in blue above each payoff type. Otherwise, if the agent does not obtain the object, her value is zero. At awareness level $\bar{\ell}$, the utilitarian ex-post efficient outcome function must allocate the object to agent 1 while at awareness level $\underline{\ell}$ it must allocate it to agent 2. The twist is that only agent 2 has awareness level $\bar{\ell}$. In the dynamic elaboration Clarke mechanisms, initially agent 1 reports $t_1'$ and agent 2 reports $t_2''$. The pooled awareness level is $\bar{\ell}$. At the next stage, agent 1 elaborates on her prior report by submitting report $t_1''$. Agent 2 has nothing to elaborate. The following table shows the computations of the $m_i$, $f_i$, and $u_i$ for $i \in \{1, 2\}$ and each awareness level. The order of the terms in each sum follows the order of terms in the sum making up the definition of $m_i$, $f_i$ and $u_i$, respectively:
\footnotesize
$$\begin{array}{|c|c|cc|cc|} \hline  & a_2 & f_1 & f_2 & u_1 & u_2 \\ \hline  
\quad \bar{\ell} \quad & \quad 0 + 3 - 0 - 1 - 2 + 2 = 2 \quad & \quad 0 - 1 - 2 = -3 \quad & \quad 2 - 2 + 2 = 2 \quad & \quad 2 - 3 = -1  \quad & \quad 0 + 2 = 2 \quad \\
\underline{\ell} & 0 & 3 - 3 + 0 = 0 & 0 - 1 + 0 = -1 & 0 + 0 = 0 & 3 - 1 = 2 \\  \hline 
\end{array}$$  
\normalsize 
We observe that for agent 1, the ex-ante anticipated ex-post participation constraint is satisfied. At his initial awareness level $\underline{\ell}$, she does not expect to obtain the good and does not expect to pay. However, upon agent 2 raising pooled awareness to $\bar{\ell}$, agent 1 realizes that she will obtain the good but beyond the Clarke payments of 1 she also has to compensate agent 2 for raising awareness with an additional payments of 2. Thus, her ex-post utility is -1, violating her ex-post participation constraints at interim after stage 1. \hfill $\Box$\\

The example underlines the importance of agents' ex-ante commitment to participating in the mechanisms in the face of unanticipated experiences.\footnote{Typically, contracts exposing agents to unanticipated experiences such as enlisting in the military also pose draconian penalties for reneging the commitment. For instance, desertion was often punishable by death, especially during wartime.}

Showing that the dynamic elaboration Clarke mechanism satisfies ex-ante anticipated ex-post participation constraints relies on the fact that agents do not consider other agents to be more aware than themselves. One feature of unawareness is that if an agent is unaware of an event, then she is unaware that she is unaware of the event (so called AU introspection; see \citet{HMS2006}). While an agent who is unaware of an event cannot contemplate about this particular event and hence cannot contemplate about other agents' awareness of this event, she could nevertheless be aware that she might be unaware of something that other agents are aware of. \citet{S2024} extended unawareness structures of \citet{HMS2006, HMS2013a} to awareness of unawareness.\footnote{For other approaches to awareness of unawareness in interactive settings, see \citet{HalpernRego2013} and \citet{BoardChung2021}. In contrast to \citet{S2024}, those approaches are not event-based.} Since our payoff type space structure is a simplified version of unawareness structures, we could also extend them to awareness of unawareness and model explicitly the state of the mind when an agent contemplates her participation constraints under awareness of unawareness. In such a model, there can be two situations, one in which the agent is unaware of her unawareness and one in which she has the same payoff type but is aware of her unawareness. While in the former, the agent would happily sign on the mechanisms, in latter the agent might be reluctant to do so because although she cannot anticipate what she is currently unaware of, she suspects that she has to pay agent $i$ for raising awareness.  Such a scenario is quite realistic as agents may have experience in prior mechanisms and are able to reason about other agents having less awareness.  We leave the extension to future work. However, in the following section we develop a mechanism for procurement that satisfies participation constraints even after awareness is raised.

\section{Procurement under ex-ante Unforeseen Contingencies\label{sec:RSPA}} 

A common mechanism used in procurement is the reverse action. Despite the similarly of the second price auction (\cite{Vickrey1961}) to the Clarke mechanism, the reverse second price auctions is not an example of the Clarke mechanism. The reasons are twofold: First, as usual in economics textbooks, we defined $f^{-i}_0$ by $f^{-i}_0(\bm{t}) = \arg \max_{x_{0} \in X_{0}} \sum_{j \neq i} v_j(x_{0}, t_j)$. With such a definition the payment received by the winning bidder uses an outcome where $i$ supplies the good when computing the maximum wlefare without $i$'s cost/value function. Thus, she does not pay the second smallest bid but zero, which differs from the reverse second price auction.\footnote{We could redefine dynamic elaboration Clarke mechanisms by letting $f^{-i}_0$ be the maximum of the sum of valuations of agent $i$'s opponents only over outcomes still available without agent $i$. In such a case, we would need to require further assumptions in order to show no deficit; see \citet{ShohamLeyton-Brown2012}.}  The other reason is that the buyer is special in that her payments are not computed using the Clarke transfers but she pays the second lowest bid to the winning bidder as required by the reverse second price auction. She is what has been called a ``sink-agent'' (e.g., \citet{NathSandholm2019}) with respect to transfers. Since procurement of complex projects under ex-ante unforeseen contingencies and incomplete specifications is an extremely relevant topic, we like to extend our setting to it. In particular, using dynamic direct elaboration mechanisms we aim to remedy a shortcoming of standard reverse second price auctions and allow the buyer to take advantage of the expertise of bidders by pooling their awareness. 

Let $I = \{1, ..., b\}$. We say that $\langle X_{0, -i}, c_i, \rangle_{i \in I \setminus \{b\}}$ is a \emph{procurement context} if for all $i \in I \setminus \{b\}$, $X_{0, -i} = X_{0, -i}^{\ell} = \{\emptyset\} \cup (I \setminus \{b, i\})$ for all $\ell \in L$, and $c_i = - v_i$ is the cost function of agent $i \in I \setminus \{b\}$. This is interpreted as follows: There is a special agent, the buyer denoted by $b$. The other agents, $I \setminus \{b\}$, are potential suppliers/sellers. The set of outcomes $X_{0, -i} = \{1, 2, i-1, i+1, ..., b-1, \emptyset\}$ specifies which of the sellers in $\{1, 2, i-1, i+1, ..., b-1\}$ supplies the good or whether nobody supplies the good, $\emptyset$. This is the set of outcomes feasible without supplier $i$. The value $c_i(t_i)$ is seller $i$'s cost of supplying the good when her type is $t_i$. If she does not supply the good, her cost is zero. Note that in this setting, the strategy of agent $i$ is still to report an entire payoff type, while $c_i(t_i)$ would be agent $i$'s total cost/value associated with the reported payoff type $t_i$. 

\begin{definition}[Dynamic Elaboration Reverse Second Price Auction]\label{def:RSPA}  Given a procurement context, the dynamic direct elaboration mechanism implementing $f$ is a dynamic elaboration reverse second price auction if $f_0$ assigns the project to a lowest bidder (with an arbitrary but ex-ante given tie breaking rule) and transfers $f_i$ to seller $i \in I \setminus \{b\}$ are given for any sequence of reported payoff type profiles $(\bm{t}^1, ..., \bm{t}^n) \in \mathfrak{T}$ by
\begin{eqnarray}\label{eqn:RSPA1} f_i(\bm{t}^1, ..., \bm{t}^n) := \mathbb{I}_i(\bm{t}^n) c_{(2)}(\bm{t}^n) + a_i(\bm{t}^1, ..., \bm{t}^n),
\end{eqnarray} where for $i \in I \setminus \{b\}$ indicator function $\mathbb{I}_i$ is defined by 
\begin{eqnarray} \mathbb{I}_i(\bm{t}^n) & := & \begin{cases} 1 & \mbox{if } f_0(\bm{t}^n) = i \\
0 & \mbox{otherwise} \end{cases} 
\end{eqnarray} and $c_{(2)}(\bm{t}^n)$ is the second smallest order statistic of $\{c_i(t_i^n)\}_{i \in I \setminus \{b\}}$. The term $a_i(\cdot)$ is defined for $i \in I \setminus \{b\}$ by
\begin{eqnarray}\label{eqn:RSPA2}
a_i(\bm{t}^1, ..., \bm{t}^n) :=
\begin{cases}
m_i(\widecheck{\lambda}(\bm{t}^n)) & \mbox{if } i = i^*(\bm{t}^1, ..., \bm{t}^n) \\
0 & \mbox{otherwise}
\end{cases}
\end{eqnarray}
where as before
\begin{eqnarray*} i^*(\bm{t}^1, ..., \bm{t}^n) & := & \left\{i \in I : \exists k \leq n \ \left((\lambda(t_i^k) = \widecheck{\lambda}(\bm{t}^n)) \mbox{ and } \not\exists j \neq i, k' \leq k \ (\lambda(t_j^{k'}) = \widecheck{\lambda}(\bm{t}^n))\right)\right\}
\end{eqnarray*} and $m_i(\ell)$ is defined recursively, as follows: $m_i(\underline{\ell}) :=  0$ and for any $\ell \triangleright \underline{\ell}$,
\begin{eqnarray}\label{eqn:RSPA3} m_i(\ell)  & := &  \max\limits_{\ell' \triangleleft \ell, \bm{t}' \in \bm{T}^{\ell'}, \bm{t} \in \bm{T}^{\ell}} \left\{\left( m_i(\ell') + \mathbb{I}_i(\bm{t}') \left(c_{(2)}(\bm{t}') -  c_i(t_j')\right) - \mathbb{I}_i(\bm{t}) \left(c_{(2)}(\bm{t}) -  c_i(t_j)\right)\right), 0\right\}.
\end{eqnarray} Finally, the transfer to the buyer is 
\begin{eqnarray} f_b(\bm{t}^1, ..., \bm{t}^n) & := & - c_{(2)}(\bm{t}^n) - \sum_{i \in I \setminus \{b\}} a_i(\bm{t}^1, ..., \bm{t}^n).
\end{eqnarray}
\end{definition}

In the dynamic elaboration reverse second price auction, there are a finite number of stages of raising awareness after which the project is awarded to the bidder with the lowest bid price. She is paid by the buyer the second lowest bid price. Any bidder, no matter whether she is awarded the contract or not, may receive a compensation for raising awareness if she is the unique agent who is the first to raise awareness to the pooled awareness level. In such a case, the agent is also compensated for the best possible profit she would lose by pretending to be less aware, which is similar to the $m_i$-terms of the transfers of the dynamic elaboration Clarke mechanisms. The buyer pays the second lowest bid to the lowest bidder and the compensation for raising awareness if any. Note that the unique agent raising the awareness level to the pooled awareness level may also be the buyer, in which case she would not have to pay any compensation for raising awareness to the sellers. She also does not pay any compensation for raising awareness if there is no unique bidder who raises awareness to the pooled awareness level or if the pooled awareness level is the join that is strictly greater than awareness levels communicated by the sellers. 

Note that if we assume that for every awareness level and seller there is a payoff type profile with which she does not have the lowest bid price or she could always opt out even if it would be utilitarian ex-post efficient for her to produce, then the definition of $m_i(\ell)$ of equation~(\ref{eqn:RSPA3}) simplifies to 
$$m_i(\ell)  = \max\limits_{\ell' \triangleleft \ell, \bm{t}' \in \bm{T}^{\ell'}} \left( m_i(\ell') + \mathbb{I}_i(\bm{t}') \left(c_{(2)}(\bm{t}') -  c_i(t_i')\right)\right).$$ 

The dynamic elaboration reverse second price auctions satisfies the following: 

\begin{theorem}\label{theo:RSPA} Given any procurement context, the dynamic elaboration reverse second price auction implements a utilitarian ex-post efficient outcome under pooled awareness in conditionally dominant strategies with budget balance and satisfies ex-post participation constraints of all sellers/bidders.
\end{theorem}

The proof of utilitarian ex-post efficiency in conditional dominant strategies under pooled awareness is analogous to the proof of Theorem~\ref{theo:VCG} and thus omitted. Budget balance follows immediately from the fact that the buyer makes any payments to the sellers. Satisfaction of ex-post participation constraints for bidders follow from the fact that at worst, a bidder is left with zero transfers, in which case she also does not supply the object. If she supplies the object, the must be the lowest bidder but is paid the second lowest bid, leaving her a non-negative profit. Note that not just ex-ante anticipated ex-post participation constraints are satisfied for each bidder but even the interim anticipated ex-post participation constraints are satisfied. Moreover, awareness of unawareness does not play any role. Even if a bidder is aware that other bidders may be aware of something that she herself is not, her participation constraints are satisfied.

\section{Discussion\label{sec:discussion}} 

One might be concerned that our mechanisms incentivize agents to raise awareness of irrelevant events in order to receive larger side payments. Formally, this will not affect any of our results, since to the extent these events are irrelevant, awareness of them will not change the physical outcome. However, we may still be concerned that this will make payments larger --- in the case of the reverse auction, payments from the mechanism. This concern can be addressed by modifying the mechanism so that it ignores awareness of irrelevant events when computing the transfers. Formalizing this modified mechanism comes at the cost of some additional notation, so we relegate it to the \href{https://faculty.econ.ucdavis.edu/faculty/schipper/appendix_mechunaw.pdf}{Online Appendix}. 

Our results appear to contradict \citet{JehielMoldovanu2001}'s impossibility theorem on efficient implementation under interdependent valuations. We achieve ex-post efficiency despite asymmetric awareness introducing interdependent values because agents can only report less awareness than they have, not more (note also we assume private values {\it given} an awareness level). This creates a unidirectional incentive compatibility environment. \citet{KraehmerStrausz2024} show that when agents can only report lower types, incentive compatibility does not restrict implementable allocation rules. Although in their model the order on types is also connected to payoffs, while in our model there is no necessary connection between awareness and payoffs, this assumption is only needed for other results in their paper. In both cases, the key insight is that sufficient transfers can incentivize truthful reporting when truth is the maximal feasible report.

It is interesting to note that \citet{Mezzetti2004} overcomes the impossibility of ex-post efficient implementation under interdependent valuations with two-stage Groves mechanisms in which agents first report benefits from outcomes and transfers are determined only after every valuation becomes transparent. Our mechanism has some similarity to his, in that transfers are determined only after payoffs are clarified and reported in a second stage. However, unlike in Mezzetti, we require a second stage of reporting to determine the allocation, and the reason we are able to elicit truthful reports is because of the ``evidence''-like properties of awareness. In contrast, Mezzetti utilizes payoff information reported at an ex-post stage where payoff interdependencies are no longer relevant. Moreover, Mezzetti's mechanism is incentive compatible in (ex-post) equilibrium rather than dominant strategies. 

While this is the first paper on general mechanism design under unawareness, it is just a first step. One shortcoming is that upon becoming aware, we assume that their more elaborated payoff type becomes immediately transparent to agents. While sometimes raising awareness may be an ``eye opener", other times it may also require substantial efforts on part of the agents to acquire information on relevant events and issues they newly became aware. In a natural next step, we study information acquisition upon becoming aware using ideas from \citet{BergemannVaelimaeki2002}.

The next step is to ask for optimal (e.g., revenue maximizing) mechanism design under unawareness. How would a designer maximize surplus from agents with heterogeneous awareness. As standard mechanism design, we would need to go beyond belief-free mechanism design. In our dynamic context, this would be complicated by updating of beliefs in light of new information and awareness. Since in standard mechanism design, optimal mechanism design is the multiagent extension of screening problems, under unawareness the screening problem studied by \citet{FrancetichSchipper2024} should be relevant. Also \citet{LiSchipper2024} study raising bidders' awareness before second price auctions in a revenue maximizing way. Nevertheless, optimal mechanism design under unawareness remains an unexplored field.


\bibliographystyle{ecta-fullname} 
\bibliography{mechunaw.bib}  

\end{document}